\documentclass[12pt]{article}
\usepackage{latexsym,graphicx,amssymb,color,amsmath}
\usepackage[width=450pt,height=675pt]{geometry}
\usepackage[bottom]{footmisc}
\usepackage[numbers,sort&compress]{natbib}
\usepackage{hyperref}
\usepackage{relsize}

\newcommand{\be}{\begin{equation}}
\newcommand{\ee}{\end{equation}}
\newcommand{\ba}{\begin{eqnarray}}
\newcommand{\ea}{\end{eqnarray}}

\begin{document}
\vspace*{-2cm}
\begin{flushright}
 DCPT/08/27
\end{flushright}
\vspace{0ex}

\begin{center}
{\larger\larger\larger {\bf DNA cages with icosahedral symmetry in bionanotechnology}}\\[2ex]
{\smaller\smaller (Article contributed to `Algorithmic Bioprocesses', A. Condon, D. Harel, J. N. Kok,  A. Salomaa, and E. Winfree Editors, Springer Verlag, 2008)}\\[-1ex]
\vspace{1cm} {\large  \bf Natasha Jonoska\,\footnote{\noindent E-mail: {\tt jonoska@math.usf.edu}}, Anne
Taormina\,\footnote{\noindent E-mail: {\tt anne.taormina@durham.ac.uk}} and Reidun Twarock\,\footnote{\noindent E-mail: {\tt rt507@york.ac.uk}}}\\

\vspace{0.3cm} {${}^{1}$}\em Department of Mathematics\\ University of South Florida,\\
Tampa, FL 33620, USA\\
 \vspace{0.3cm} {${}^{2}$\em \it Department of Mathematical
Sciences\\ University of Durham,\\ Durham DH1 3LE, U.K.}\\ 
 \vspace{0.3cm} {${}^{2}$\em \it Department of Mathematics and Department of 
 Biology,\\ University of York, \\York YO10 5DD, U.K. \\}
\end{center}
\medskip

\begin{abstract}
Blueprints of polyhedral cages with icosahedral symmetry made of circular DNA molecules are provided. The basic rule is that every edge of the cage is met twice in opposite directions by the DNA strand, and vertex junctions 
are realised by a set of admissible junction types.  As nanocontainers for 
cargo storage and delivery, the icosidodecahedral cages are of special interest as they have the largest volume per surface ratio of all cages discussed here.
\end{abstract}

\section{Introduction}
Recent advances in biotechnology provide the necessary tools to engineer cage structures from nucleic acids, and open novel avenues for applications in nanotechnology. Cages with crystallographic symmetry have already been realised experimentally in the shape of a cube \cite{cube}, a tetrahedron \cite{tetrahedron}, an octahedron \cite{octahedron} or a truncated octahedron \cite{truncatedoctahedron}, and one natural idea is to use such cages for cargo delivery or storage \cite{drug}. Moreover, models for two realisations of a cage with a non-crystallographic symmetry have recently been proposed from a theoretical point of view \cite{GTT,JT}. These studies were motivated by the hope that such mathematical considerations on the organisation of  RNA or DNA  in cages with icosahedral symmetry will aid the design of artificial cages inspired by nature. 

\begin{figure}[ht]
\begin{center}
(a)\includegraphics[width=2.0cm,keepaspectratio]{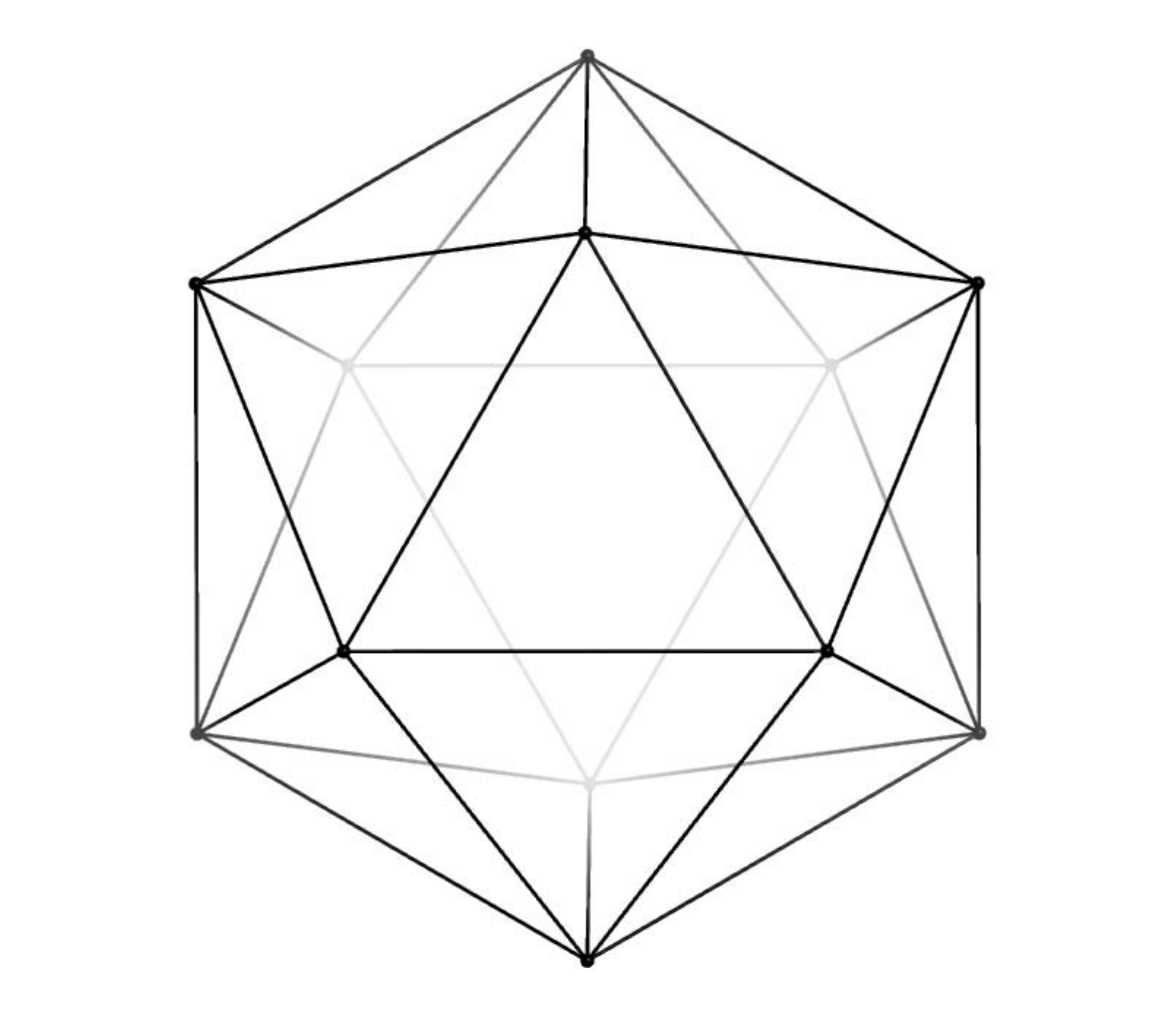}
(b)\includegraphics[width=2.0cm,keepaspectratio]{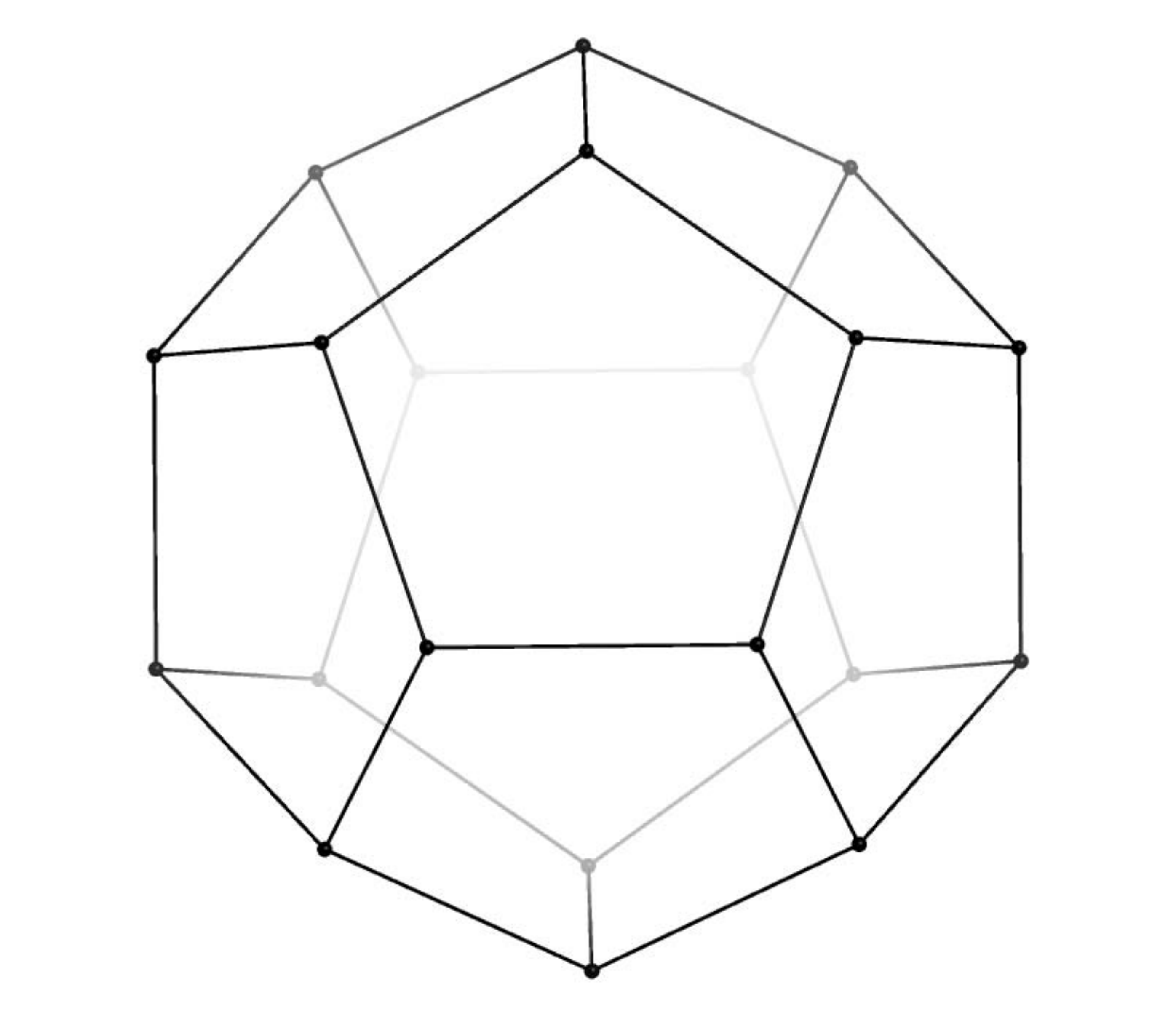}
(c)\includegraphics[width=2.0cm,keepaspectratio]{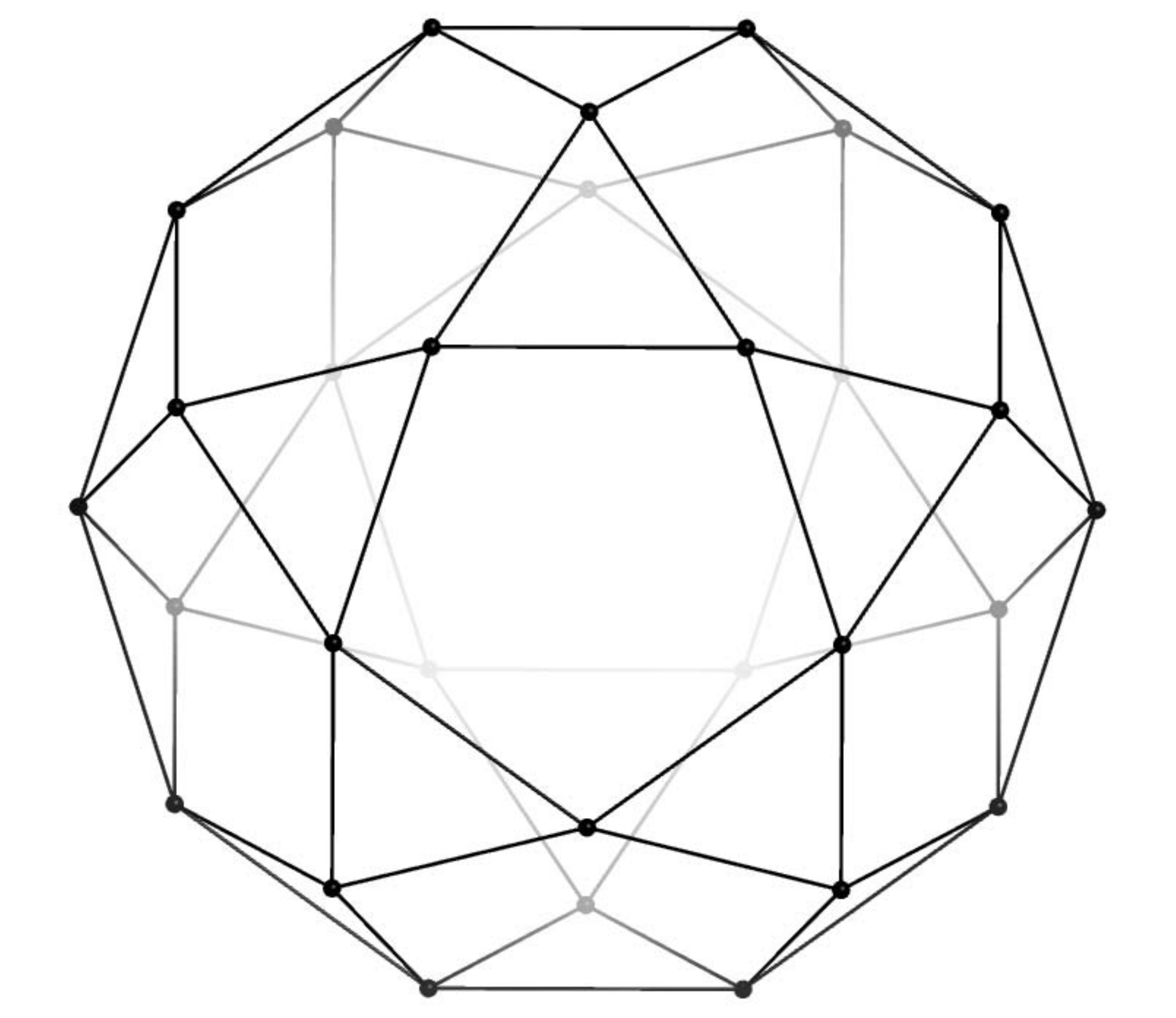}
\end{center}
\caption{{\em The icosahedron (a), dodecahedron (b) and icosidodecahedron (c) corresponding to the three polyhedra with icosahedral symmetry with less than 60 vertices.}}
\label{Polyhedra}
\end{figure}

We provide here a systematic comparative analysis of three polyhedral cages with icosahedral symmetry, the icosahedron, the dodecahedron \cite{JT} and the icosidodecahedron \cite{GTT}(see Fig. \ref{Polyhedra}), which are -- from a mathematical point of view -- distinguished because they are the three smallest vertex sets realising icosahedral symmetry. 
Moreover, their edges are uniformly of the same length. We expect that these properties make them easier to be realised  experimentally than other polyhedra with this symmetry, and we therefore focus on these three cases here. 

An interesting feature of these polyhedra is the fact that they have vertices of different connectivity: The dodecahedron has trivalent vertices and hence needs to be realised in terms of three-junctions, whilst the icosidodecahedron requires four-junctions and the icosahedron five-junctions. We investigate here possibilities of realising these polyhedra with a circular DNA molecule. Our analysis shows that the complexity of this problem increases when the
polyhedra considered have even degree vertices. The strategy implemented here is to enumerate all possible inequivalent realisations of these cages with a single strand of DNA molecule, given a set of rules on the junctions. This combinatorial approach is manageable without computer  help for the icosahedral and the dodecahedral cages, but becomes cumbersome for the icosidodecahedron.
Our treatment of the latter in this paper is therefore computer-aided, but a more elegant mathematical framework to solve this type of problem is being developed. 

Our analysis shows that in all three cases at least two circular DNA strands are needed to form the cage. From that point of view, all three polyhedra lend themselves equally well for templates of DNA cages. However, the icosidodecahedron is the polyhedron among the three with the largest volume to surface ratio (at an edge length of 1, it is approximately 0.47, compared to 0.37 for the dodecahedron, 0.25 for the icosahedron\footnote{Moreover, the volume to area ratios of all icosahedral cages considered here are larger than those of the crystallographic cages realised to date. For comparison, the value for the octahedron is approximately 0.14.}), making it perhaps the most interesting icosahedral cage for applications in nanotechnology. 

We start by introducing our theoretical construction method in general terms for all polyhedra
with icosahedral symmetry in Section \ref{sec1}, and then provide details for the three polyhedra in the subsequent sections: the icosahedral cage in Section \ref{I}, the dodecahedral cage in Section \ref{D}, and the icosidodecahedral cage in Section \ref{IDD}.

\section{Construction of cages with icosahedral symmetry: general principles}\label{sec1}

The goal is to provide blueprints of polyhedral cages with icosahedral symmetry, made of a circular DNA molecule, with the basic rule that every edge of a given polyhedral cage is met twice in opposite directions by the strand. This requirement enables hybridization of the two portions of the strand running along an edge into a double helix structure. 

Mathematically, a cage is a graph whose nodes are the vertices of the corresponding polyhedron, and the connectors are the edges.  As explained in \cite{Jonoska,Greenberg}, the idea is to topologically embed graphs into orientable thickened graphs as deformation retracts. Such thickened graphs are compact orientable 
2-dimensional surfaces constructed out of strips and thickened $n$-junctions glued together. They can be considered as models for DNA cages as follows: The boundary curves of the thickened graph represent single stranded DNA molecules in the cage, and provide a blueprint that specifies which types of junctions have to be used to realise this graph. 

The design of such templates for DNA cages must take the following factors into account:
\begin{enumerate}
\item {\em Initial data}: Assume that the cages correspond to polyhedra with all edges of equal length $\lambda$. Then   the number $\nu(\lambda)$ of half-turns in the duplex structure along each edge depends on $\lambda$. Configurations where $\nu(\lambda)$ is odd are  modelled as cross-overs in the planar projective views of the polyhedral cages as shown in Fig.~\ref{options}. 
\begin{figure}[ht]
\begin{center}
(a) \raisebox{0.6cm}{\includegraphics[width=3cm,keepaspectratio]{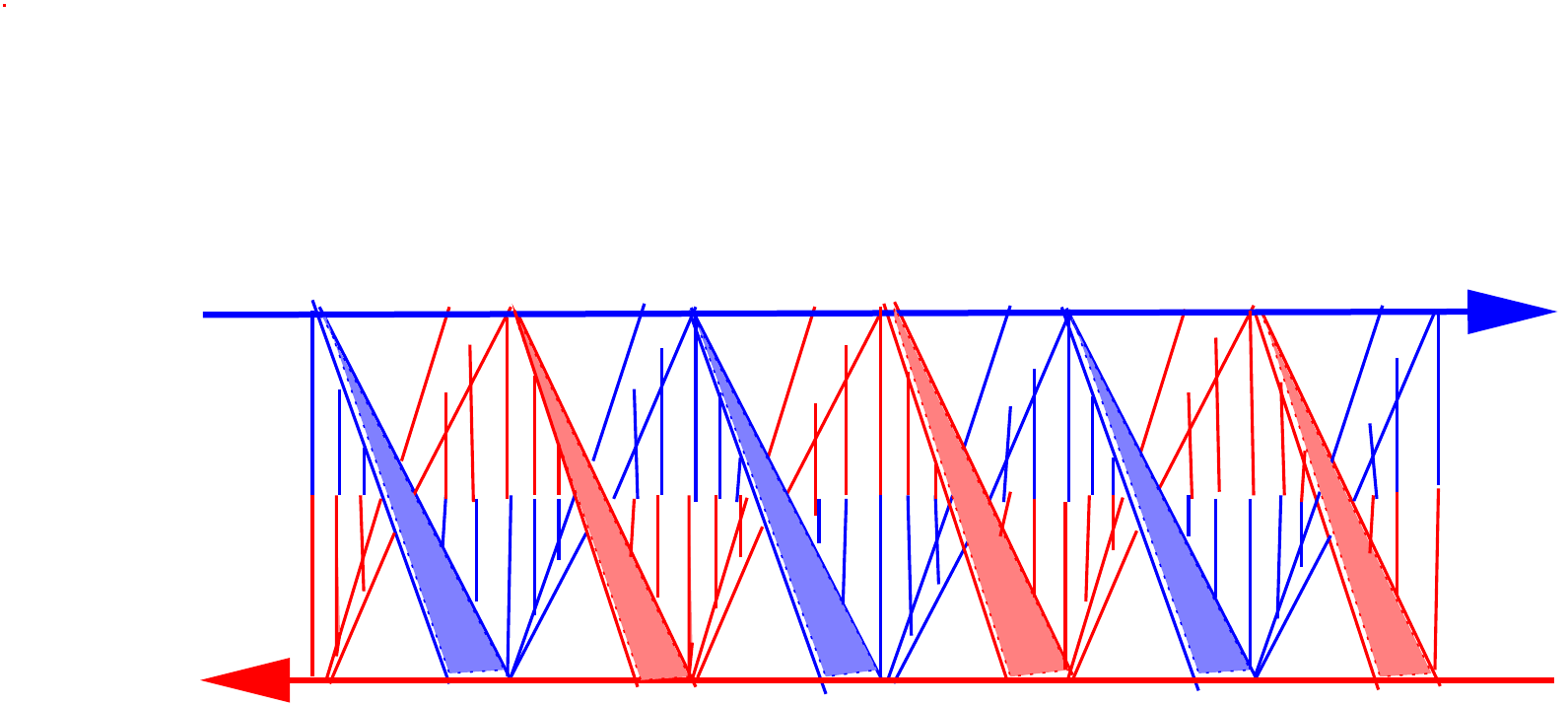}}
\raisebox{0.8cm}{$\longrightarrow$}\quad\includegraphics[width=3cm,keepaspectratio]{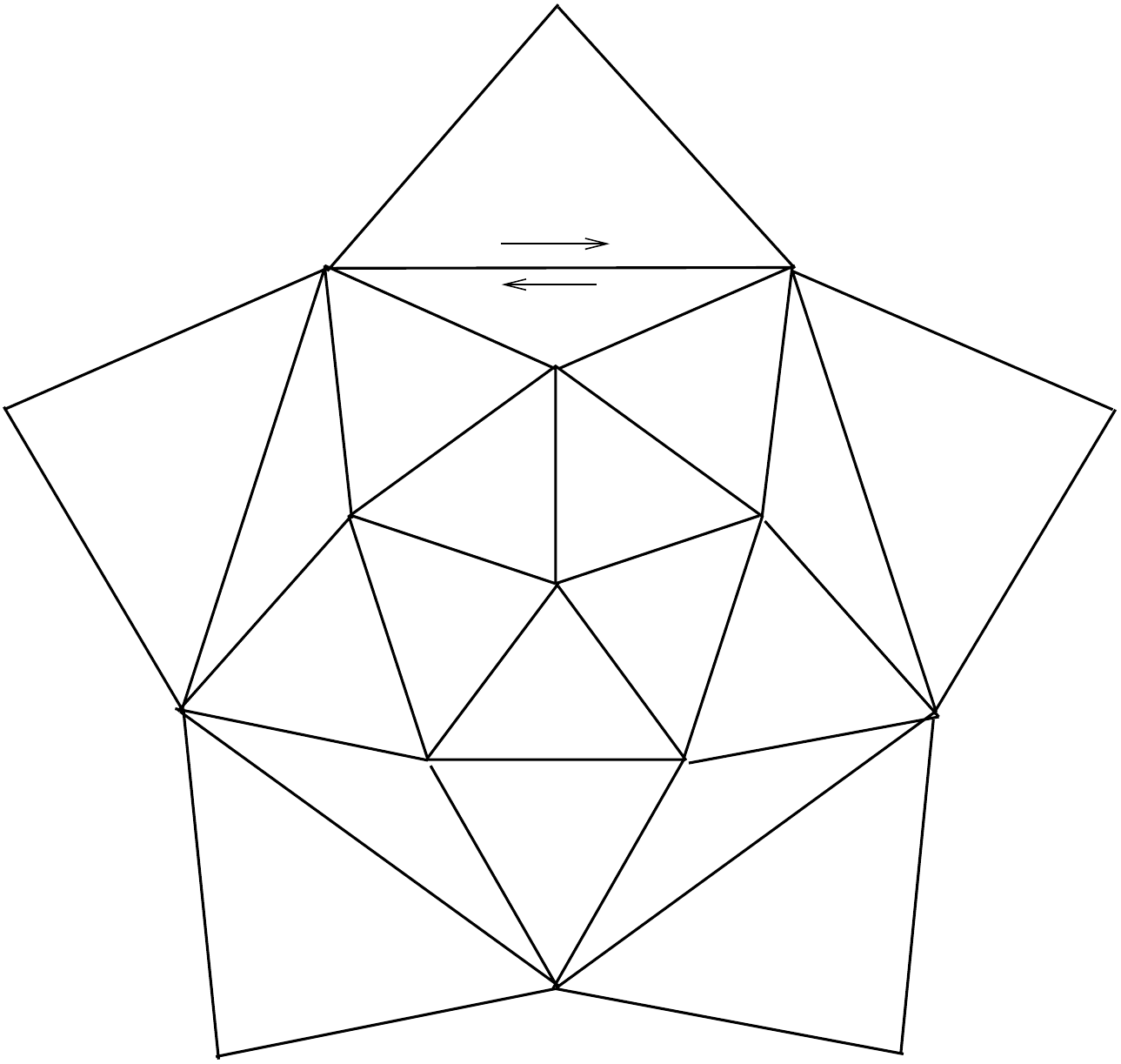}\\
(b) \raisebox{0.6cm}{\includegraphics[width=3cm,keepaspectratio]{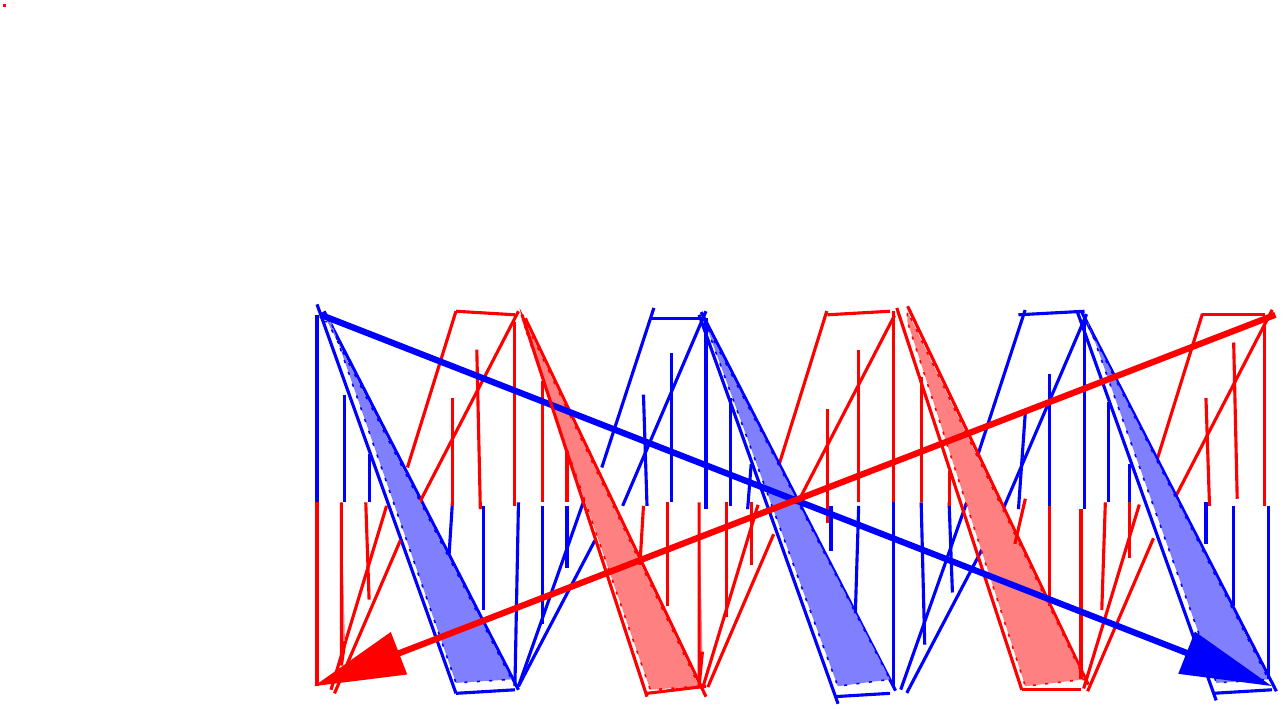}}
\raisebox{0.8cm}{$\longrightarrow$}\quad\includegraphics[width=3cm,keepaspectratio]{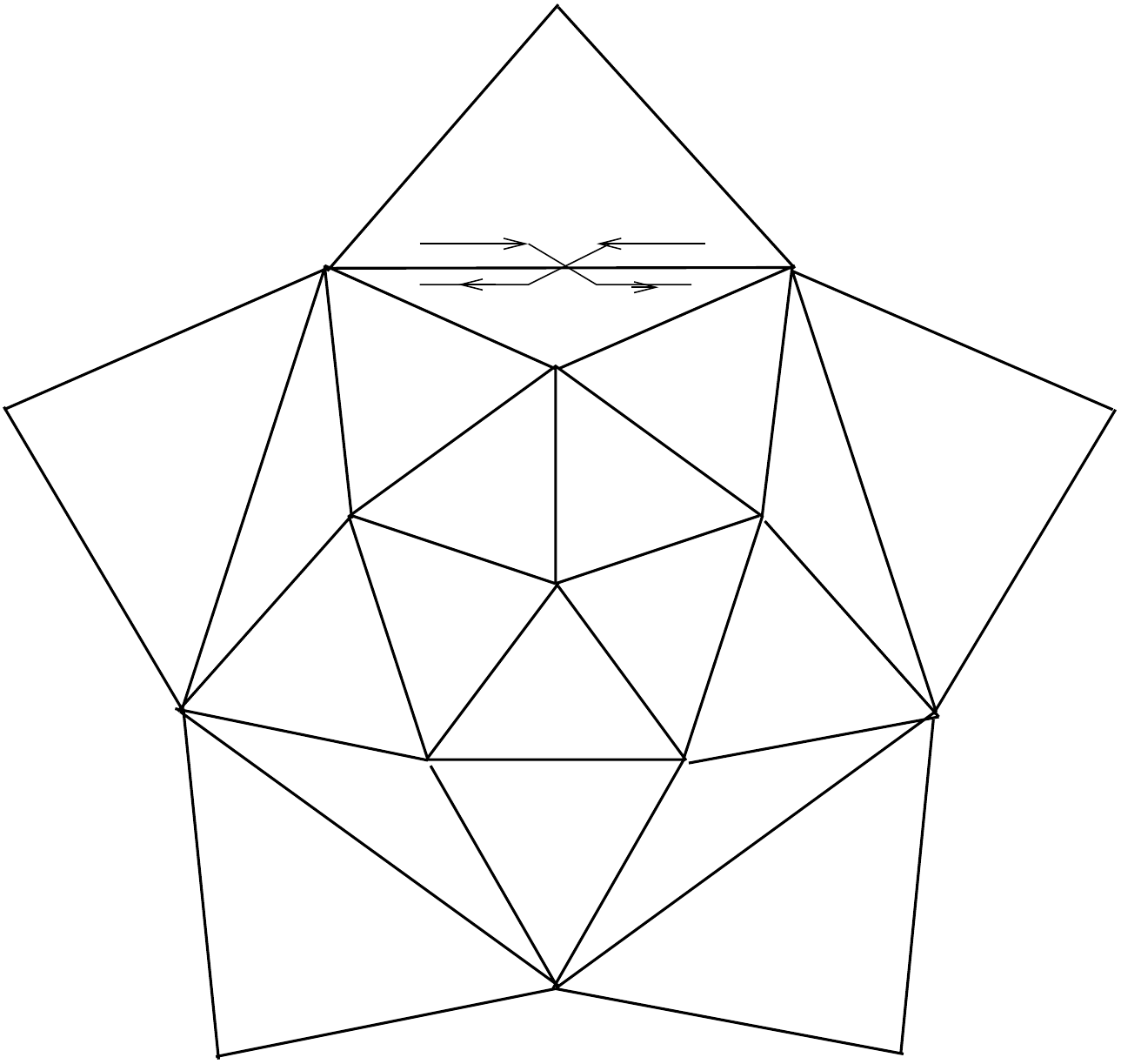}
\end{center}
\caption{{\em (a) The DNA double helix is represented by lines (blue and red) that trace the backbone of the helices, (b) depending on their lengths, an additional half-turn may appear, that is represented by cross-overs on the planar representation of the graph.} }
\label{options}
\end{figure}
Note that for DNA there are about 10.5 base pairs (bp) per helical turn.  

\item {\em Thickened $n$-junctions}: Mechanical stress may be imposed on the overall configuration if the strands of the edges cross each other at the incident vertex junctions. For example, the thickened $n$-junction shown in Fig.~\ref{OptimalVertex}(a) imposes no stress on the configuration (we name it `type A$_n$'), whilst the thickened $n$-junctions appearing in Fig.~\ref{OptimalVertex}(b) and (c) accommodate one or two cross-overs (we name them `type B$_{1n}$' and  `type B$_{2n}$') and may impose stress on the overall configuration unless extra nucleotides are introduced along the corresponding edge that compensate for it. In general, a thickened $n$-junction with $k$ cross-overs of the strands is of type B$_{kn}, k \le n$.
\end{enumerate}
\begin{figure}[ht]
\begin{center}
\raisebox{-1.3cm}{\includegraphics[width=8cm,keepaspectratio]{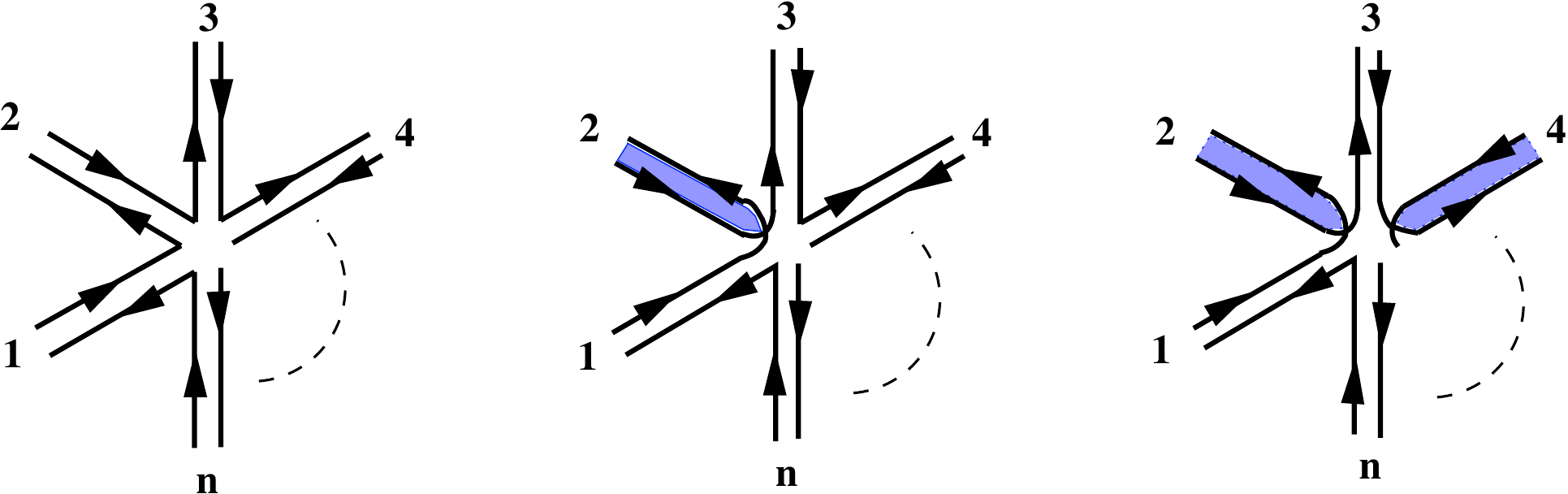}}
\end{center}
\caption{{\em (a) Energetically optimal thickened n-junction (type A$_n$); (b) higher energy  (one strand ``reaches'' across to another) thickened n-junction (type B$_{1n}$); (c) higher energy  (two strands cross-over their neighbours) thickened n-junction (type B$_{2n}$). The dotted line indicates that there may be more edges present.}}
\label{OptimalVertex}
\end{figure}
The number $n$ of legs in the junctions depends on the type of cage considered. In subsequent sections,  cages with thickened 5-junctions (icosahedra), 4-junctions \footnote{Stable four-junctions can be assembled, see for example \cite{Holliday}.}  (icosidodecahedra) and 3-junctions (dodecahedra) are discussed.

\subsection*{The construction procedure: Step I}

The first step in the construction procedure is to identify {\em start configurations}, i.e., orientable thickened graphs with a maximum number of type A$_n$ thickened junctions. Such graphs are usually made of several distinct circular strands. In order to determine the start configuration, first assume that every vertex on the polyhedron is represented by a junction of type A$_n$. This is always possible if the polyhedral edges have an even number of helical half-turns. The start configuration in this case is given by $N$ separate circular strands (loops), where $N$ is the number of polyhedral faces. 

\begin{figure}[ht]
\begin{center}
\raisebox{-1.3cm}{\includegraphics[width=5cm,keepaspectratio]{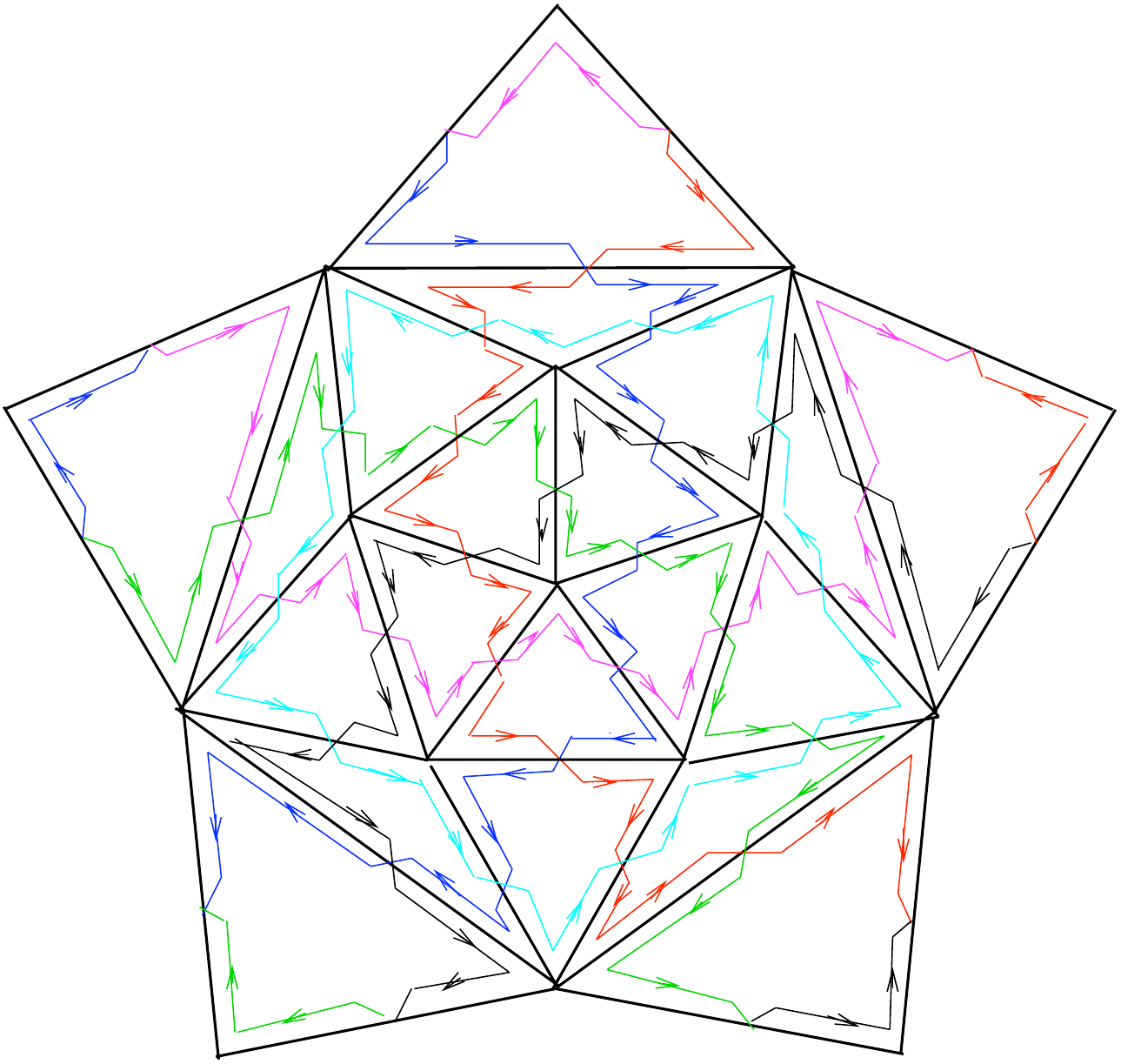}}
\end{center}
\caption{{\em Initial data configuration for an icosahedral cage when $\nu(\lambda)$ is odd. }}
\label{initialdata}
\end{figure}

\begin{figure}[ht]
\begin{center}
\raisebox{-1.3cm}{\includegraphics[width=7cm,keepaspectratio]{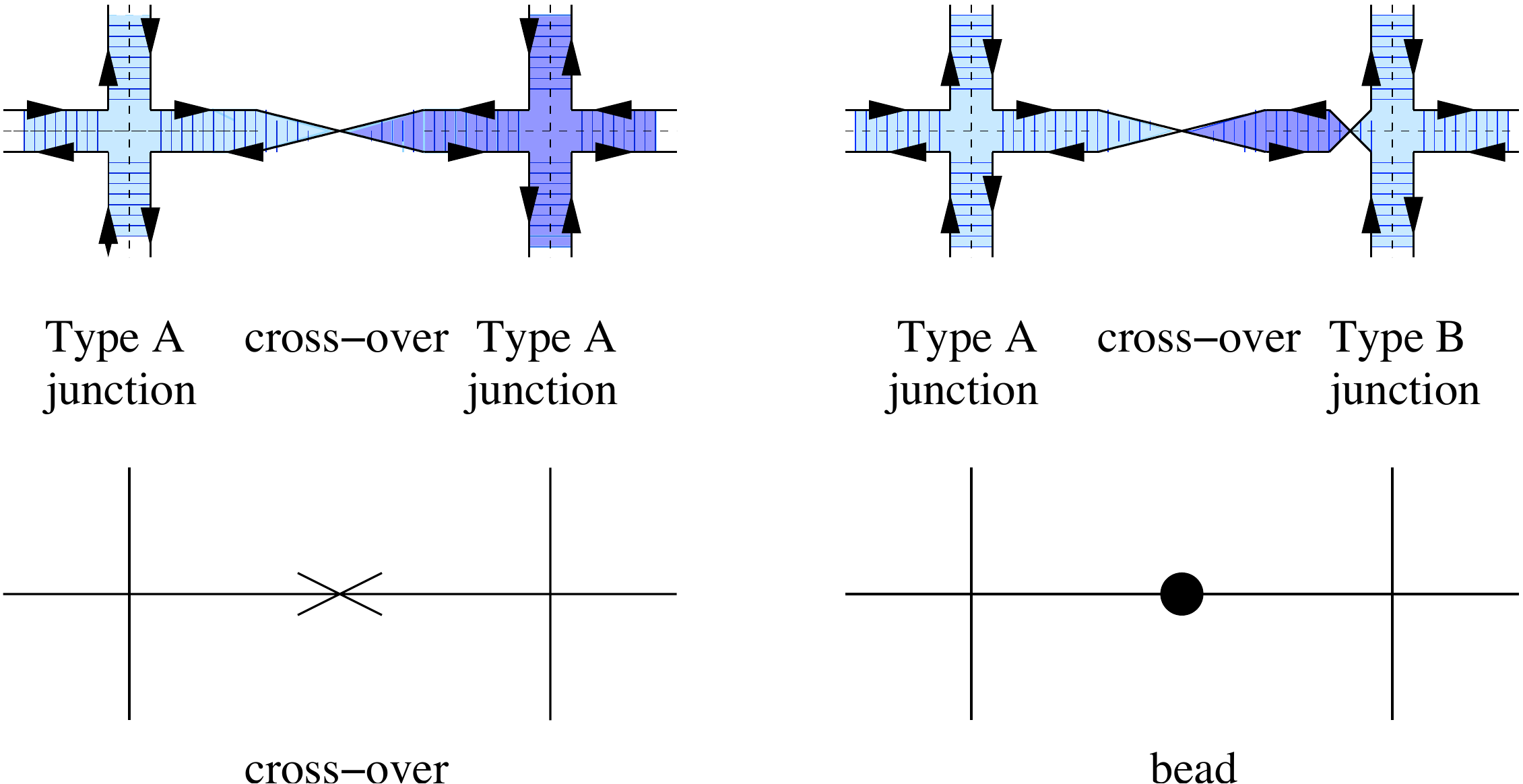}}
\end{center}
\caption{{\em Correspondence of a ``cross-over'' and a ``bead'' to the boundary orientation. An odd number of helical half-turns corresponds to a ``cross-over'' (left), and the addition of one half-turn corresponds to placing a ``bead'' on the edge.}}
\label{bead}
\end{figure}

However, in the presence of cross-overs which take into account the odd  number of half-turns along the edges, this distribution of type A$_n$ junctions 
does not necessarily provide  an orientable thickened graph. In particular, this is the case if faces with an odd number of edges occur in the polyhedron (as is the case for all three polyhedra considered here). In particular, if the cages have all edges of the same length as in the present analysis, and moreover, exhibit an odd number of half-turns in the double helix along each edge, the start configuration is obtained as follows: The two-dimensional surface, orientable or not, obtained by gluing the twisted strips representing the cross-overs to type A$_n$ junctions, is called the {\em initial data configuration} (see Fig.~\ref{initialdata} for the initial data configuration of the icosahedron; note that this configuration has 6 loops, but they do not all run in opposite directions).
To this initial start configuration, we apply the {\it bead rule} to obtain a start configuration. The bead rule consists of placing beads on selected edges of the polyhedron to indicate that a twisted strip (cross-over) is glued to a twisted leg of a type B$_{kn}$ thickened junction, as illustrated in Fig.~\ref{bead} for a 4-coordinated polyhedron.
\vskip .5cm
Note that by addition of a ``cross-over'' at one vertex the orientation of the boundary components change at all neighbouring edges. 
For polyhedra with an odd number of half-turns on their edges a start configuration is obtained if the following rules are satisfied: 
\begin{itemize}
\item Each edge accommodates either a cross-over or a bead.
\item Every face of the polyhedron in the start configuration must have an {\it even} number of cross-overs.
\item The number of beads is minimal.
\end{itemize}
After the bead rule has been applied, one must discard configurations which are equivalent under icosahedral symmetry. The symmetry-inequivalent bead configurations correspond to the possible start configurations. In such configurations some junctions are of type A$_n$, and others of type B$_{kn}$, the latter having been introduced to provide orientability of the two-dimensional surface representing the thickened graph obtained from the polyhedron. 

\subsection*{The construction procedure: Step II} 

Start configurations are usually given in terms of more than one strand of DNA. Since focus is here on building cages out of a minimal number of strands, these need to be ``merged''. This can be achieved by replacing some of the junctions in the start configuration by  junctions with different connectivity of the strands. Since such replacements result in a change of the overall number of strands used to form the cage and therefore depend on the overall structure, which is different in all three cases considered here, these replacements are discussed on a case by case basis in the following sections. 

Note that these replacements result in a change of the number of   half-turns of the DNA strands, because they add or remove a half-turn in the duplex structure \cite{JT}. If 
  the edges of the polyhedral cage are of equal length and are rigid duplexes, i.e., have hybridization going up to the vertices, then there is little 
   flexibility of the strands at the vertices to cross over from one edge to another, and  
  we assume that such replacements of the vertex structure are difficult. The 
  replacements of one type of junction with another are therefore carried out under the 
assumption that the strands at the vertices of the cage are not completely hybridized or the lengths of the edges may differ by a few nucleotides. In these cases, the changes in the 
connection of the strands introduced by these replacements do not interfere with the embedding of the double helical structure. 

\section{The icosahedral cage}\label{I}

The rules  described above can be applied in a straightforward manner to the construction of icosahedral cages, and we start by considering the case where an extra twist occurs on each edge, i.e., the edges are of equal length with an odd number of half-turns. An icosahedron is a five-coordinated polyhedron with twenty triangular faces. According to the preceding section, orientability requires an odd number of beads per face, and the minimum number of 10 beads is achieved with exactly one bead per face. 

Compiling an exhaustive list of start configurations in this case is effortless. For instance, the bead rule can be implemented by first considering one of the 12 five-coordinated vertices of the icosahedron,
 and asking  how many allowed possibilities there are to place beads on the triangular faces forming a pentagon whose centre is the vertex in question. 
 
\begin{figure}[ht]
\begin{center}
\raisebox{-1.3cm}{\includegraphics[width=8.1cm,keepaspectratio]{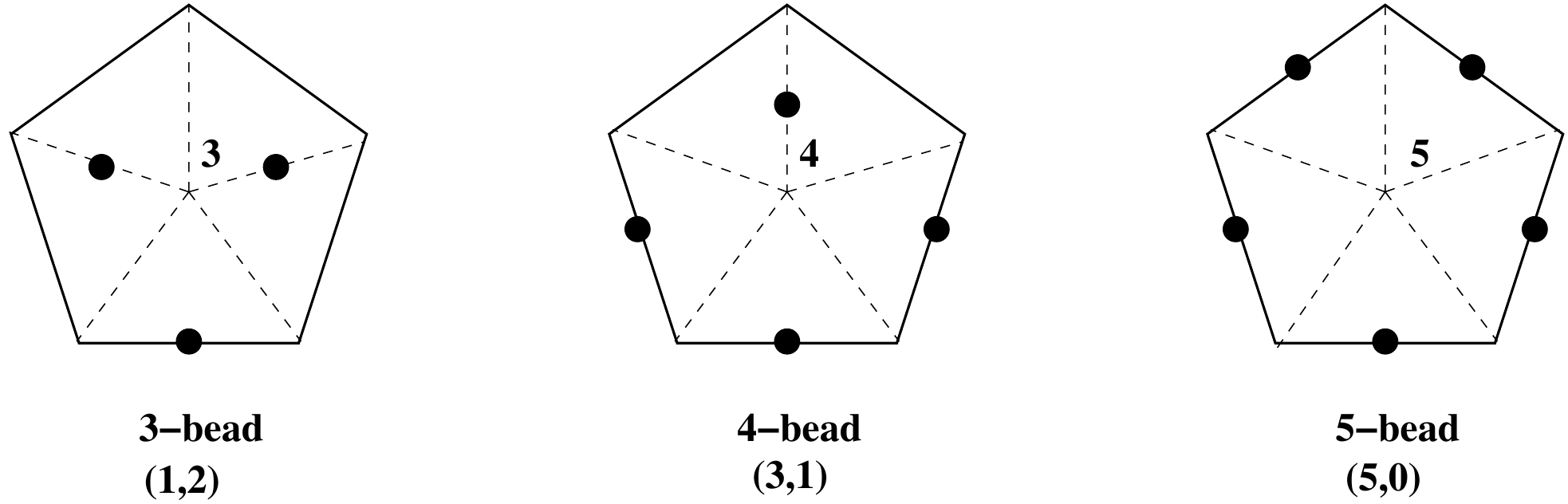}}
\end{center}
\caption{{\em Minimal bead configurations on a triangulated pentagon compatible with the bead rule.}}
\label{pentagons} \end{figure}

 Direct inspection shows that there are three inequivalent ways to obtain a minimal bead
distribution, all illustrated in Fig.~\ref{pentagons}. The 3, 4 and 5-bead configurations are labelled $(p,r) = (1,2), (3, 1)$ and $(5,0)$ respectively, where $p$ is the number of beads on the pentagon perimetre, and $r$ is the number of beads on the radii.

The following strategy provides a full set of start configurations (all letters used for vertices refer to the labelling convention shown in Fig.~\ref{vertices}(a)):
\begin{enumerate}
\item Place a $(5,0)$ pentagon configuration at any vertex (choose A, say, and label this configuration $(5, 0)_A$). Due  to the high symmetry of this bead distribution, the only bead distribution for pentagons with center vertices B,C,D,E and F is the 3-bead one, as the corresponding vertices automatically have at least two beads on radii. There is no allowed bead distribution with more than two beads on radii, so $(1,2)$ is the only possibility. A similar argument holds for the vertices G, H, I, J and K, which must be of type $(1, 2)$. Finally, vertex L is automatically of type $(5,0)$. The essential features of this icosahedral bead  configuration are encoded in Fig.~\ref{vertices}(b), where each vertex is labelled by the number of beads placed on its incident edges. This start configuration (without further replacements of $5$-junctions) requires four DNA single strands.
\begin{figure}[ht]
\begin{center}
(a)\,\,\includegraphics[width=3.5cm,keepaspectratio]{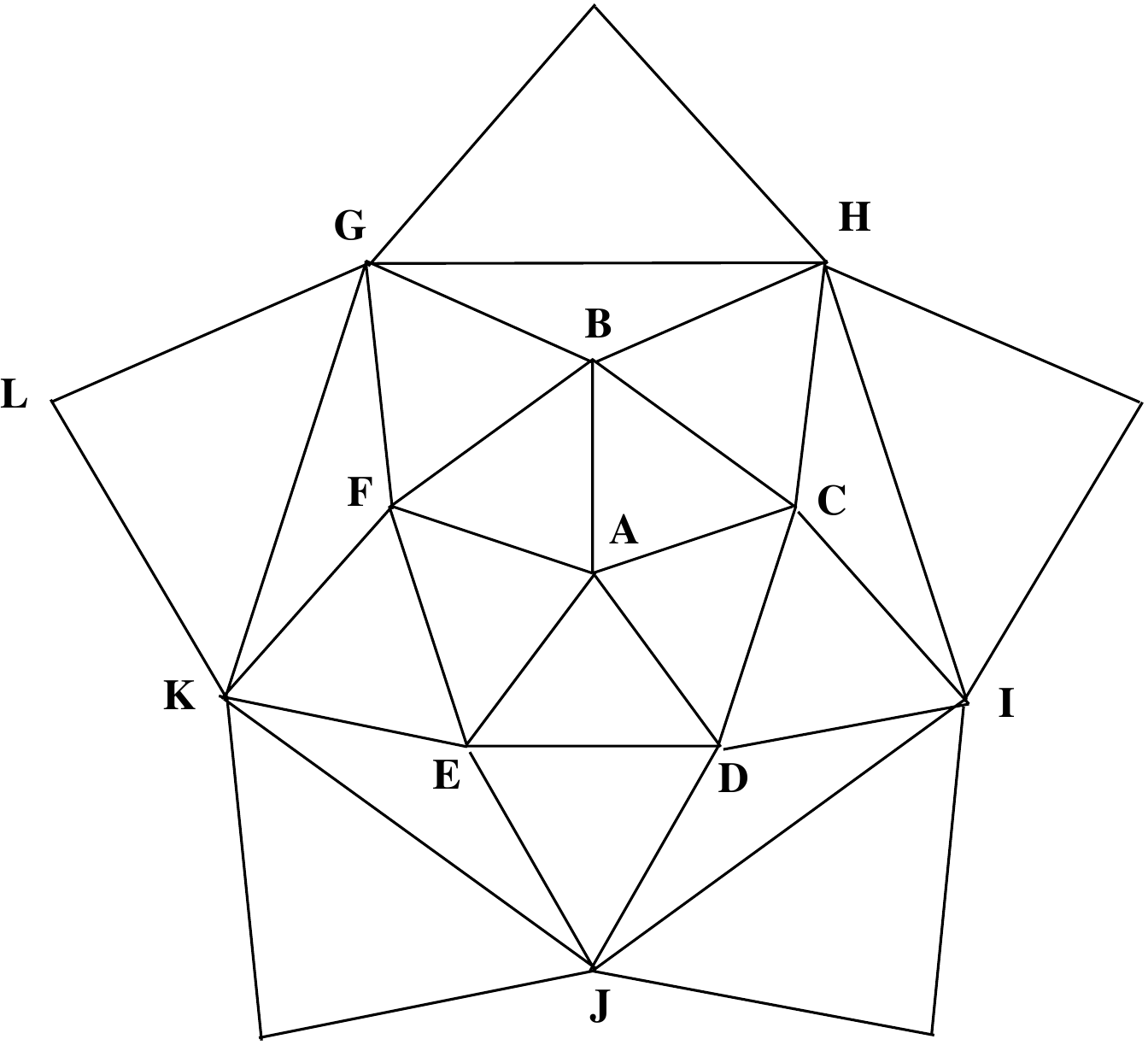}\qquad
(b)\,\,\includegraphics[width=3.5cm,keepaspectratio]{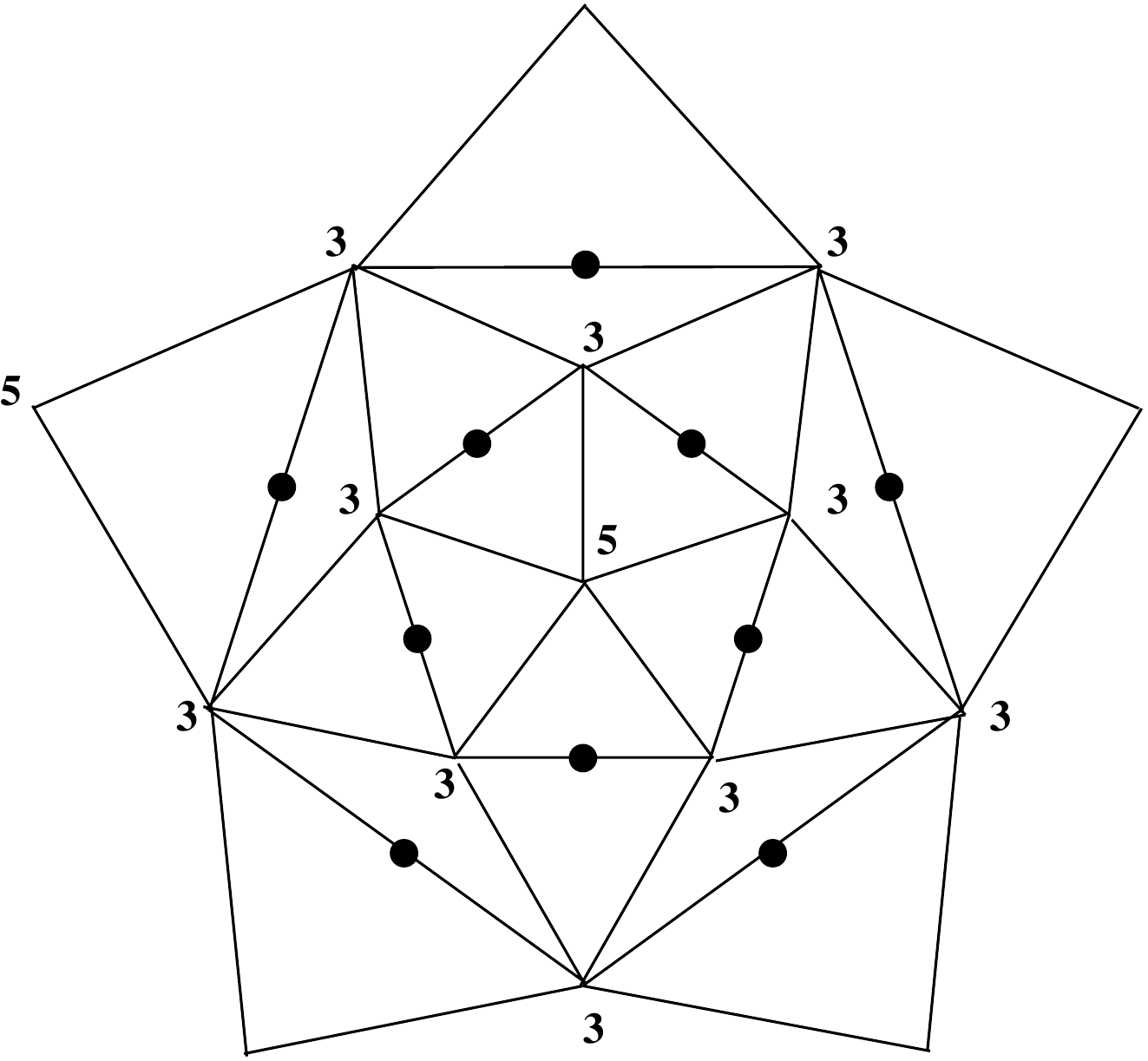}\\
\end{center}
\caption{{\em (a)  Choice of vertex labels on the icosahedron, (b) start configuration when the bead distribution $(5,0)$ is placed at vertex A (configuration 1 in Table~\ref{configs}). The number associated to each vertex counts the number of beads in the  pentagon whose centre is that vertex. }}
\label{vertices}
\end{figure}
\item Place a $(3,1)$ pentagon configuration at any vertex in any orientation. Choose vertex A, say, and place this $(3,1)_A$ configuration such that the edge linking vertices A and B hosts a bead. This fixes vertices D, E and J to have  $(1, 2)$ configurations. Vertex B can be in a $(3, 1)$ or a $(1, 2)$ configuration. The former yields that that the minimal number bead distribution makes the corresponding thickened graph not orientable, as one triangular face acquires two beads. The configuration $(1, 2)$ can be placed on vertex B in two orientations: the `left' orientation $(1, 2)_{\ell, B}$ shows a bead on the edge linking B and G, while the `right' orientation $(1, 2)_{r, B}$ shows a bead on the edge linking B and H. 
Note that B lies on the axis connecting A, B and J and a bead on the edge connecting B and G is a reflection of the bead placed on the edge connecting B and H, therefore $(1, 2)_{l,B}$ and $(1, 2)_{r,B}$ yield equivalent bead configurations (see Fig.~\ref{4}). 
\begin{figure}[ht]
\begin{center}
\raisebox{-1.3cm}{\includegraphics[width=7.5cm,keepaspectratio]{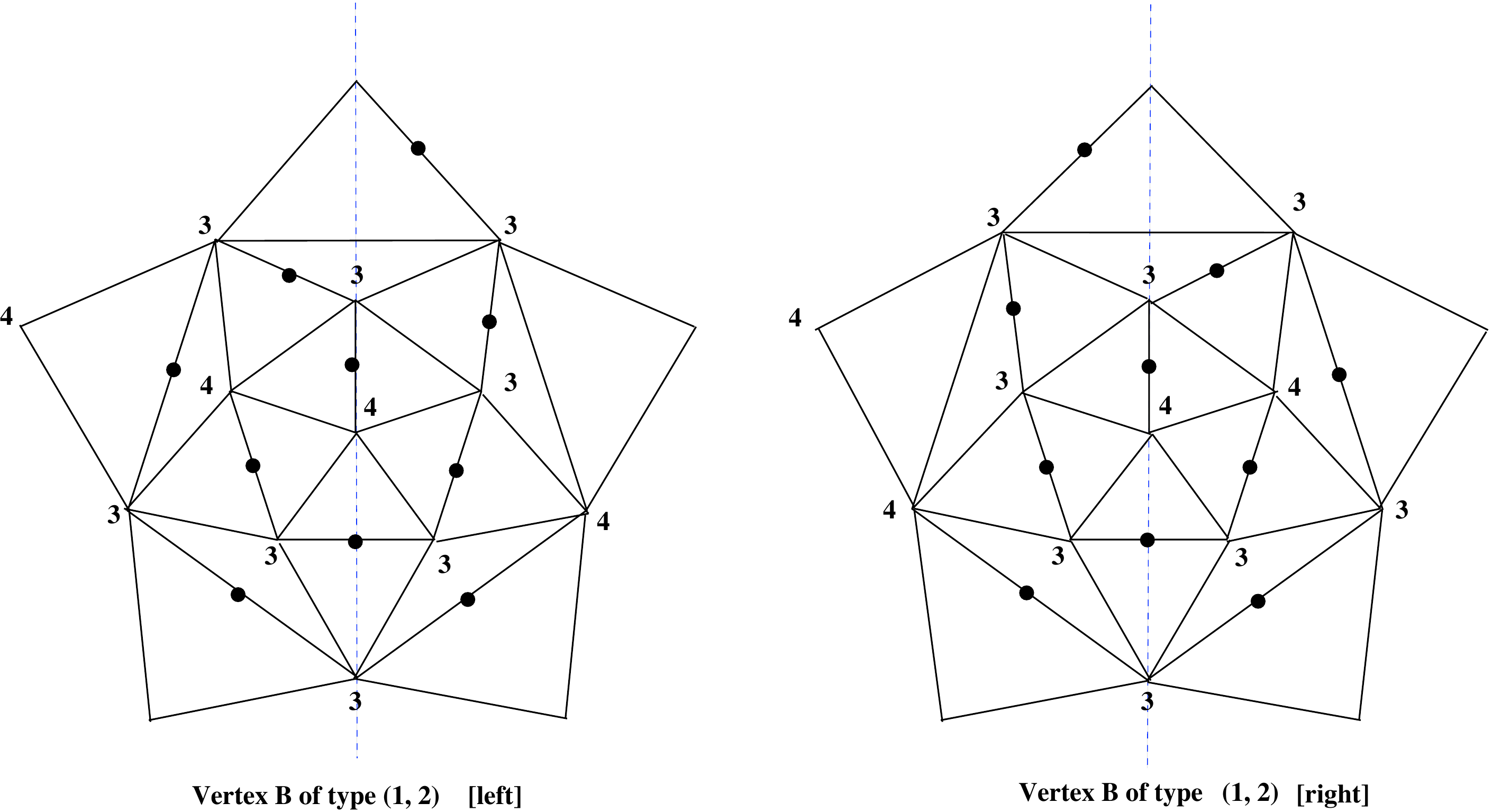}}
\end{center}
\caption{{\em Mirrored start configurations when the bead distribution $(3, 1)$ is placed at vertex A (configurations 3 and 4 in Table~\ref{configs}). The axis of mirror symmetry is the dashed blue line through vertices A, B, J and L.}}
\label{4} \end{figure}
\item  Place a $(1, 2)$ pentagon configuration at any vertex in any orientation. Choose vertex A, say, and place this $(1, 2)_A$ so that the edge linking vertices D and E hosts a bead. Then vertex B must be in a $(5, 0)$ or a $(3, 1)$ configuration since two edges on the corresponding pentagon perimetre host beads. If B is of type $(5, 0)$, one arrives at the same configuration as discussed in case 1 (Fig.~\ref{vertices} (b)). If B is $(3, 1)$, the configuration can be placed in two orientations: $(3, 1)_{\ell, B}$ if the edge linking B and G hosts a bead, and $(3, 1)_{r, B}$ if the edge linking B and H hosts a bead. Once the orientation is chosen, there is one more choice to make. In the `left' configuration, vertices C and D are automatically  in $(1, 2)$ configurations but vertex F can be of type $(1, 2)$ or $(3, 1)$. Each choice leads to a start configuration which requires two DNA strands.   In the `right' configuration, vertices E and F are automatically  in $(1, 2)$ configurations but vertex C can be of type $(1, 2)$ or $(3, 1)$. Each choice once again leads to a start configuration which requires two DNA strands.  These configurations are shown in Fig.~\ref{3a} and  Fig.~\ref{3b}. Note that two of these four start configurations are mirror images of the other two.
\end{enumerate}
\begin{figure}[ht]
\begin{center}
\raisebox{-1.3cm}{\includegraphics[width=7.5cm,keepaspectratio]{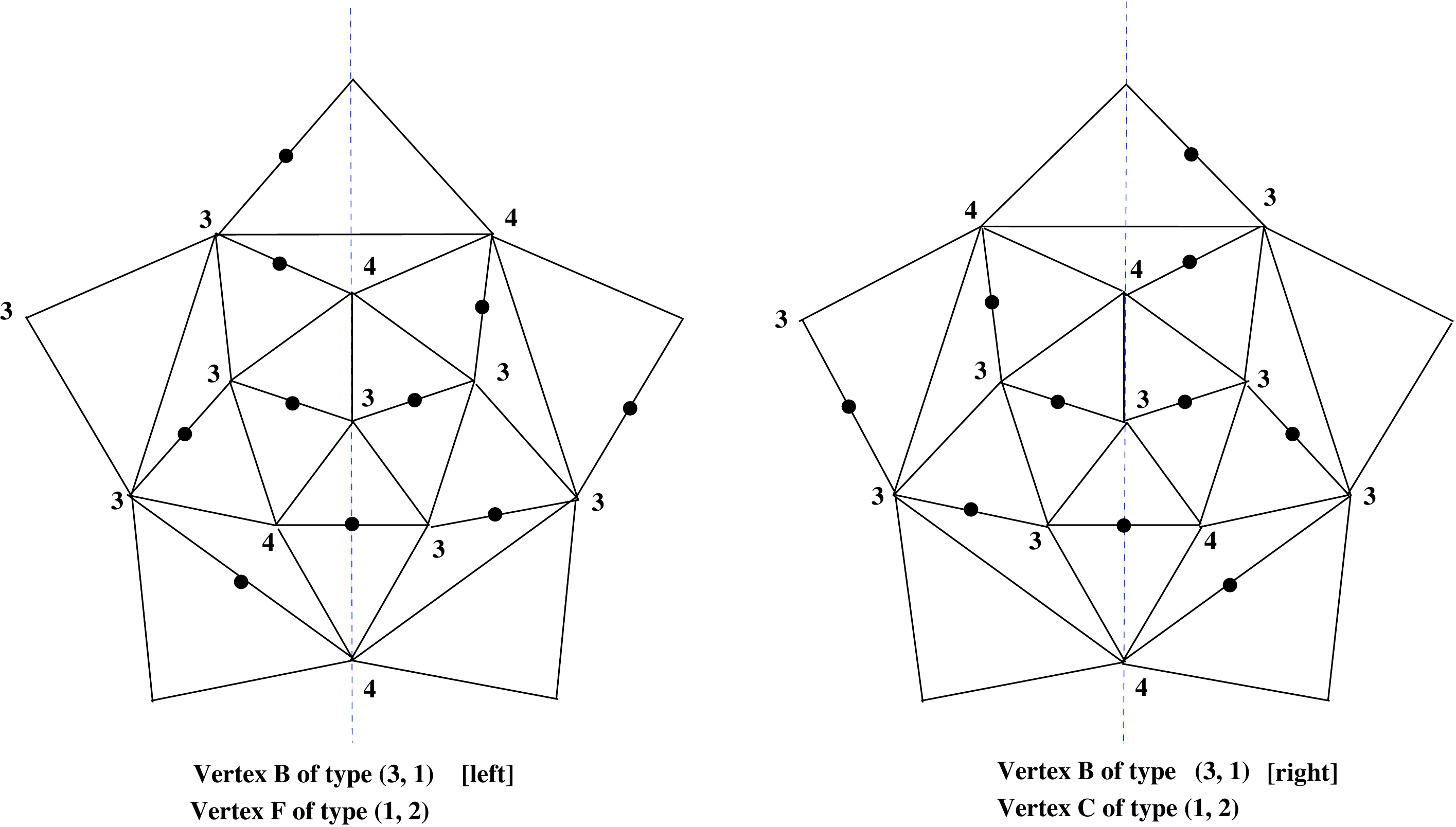}}
\end{center}
\caption{{\em Mirrored start configurations when the bead distribution $(1,2)$ is placed at vertex A, with either $(3, 1)_{\ell, B}$ and $(1, 2)_F$  (configuration 6 in Table~\ref{configs}), or with 
 $(3, 1)_{r, B}$ and $(1, 2)_C$ (configuration 8 in Table~\ref{configs}). The axis of mirror symmetry is the dashed blue line through vertices A, B, J and L.}}
\label{3a}\end{figure}
 
\begin{figure}[ht]
\begin{center}
\raisebox{-1.3cm}{\includegraphics[width=7.5cm,keepaspectratio]{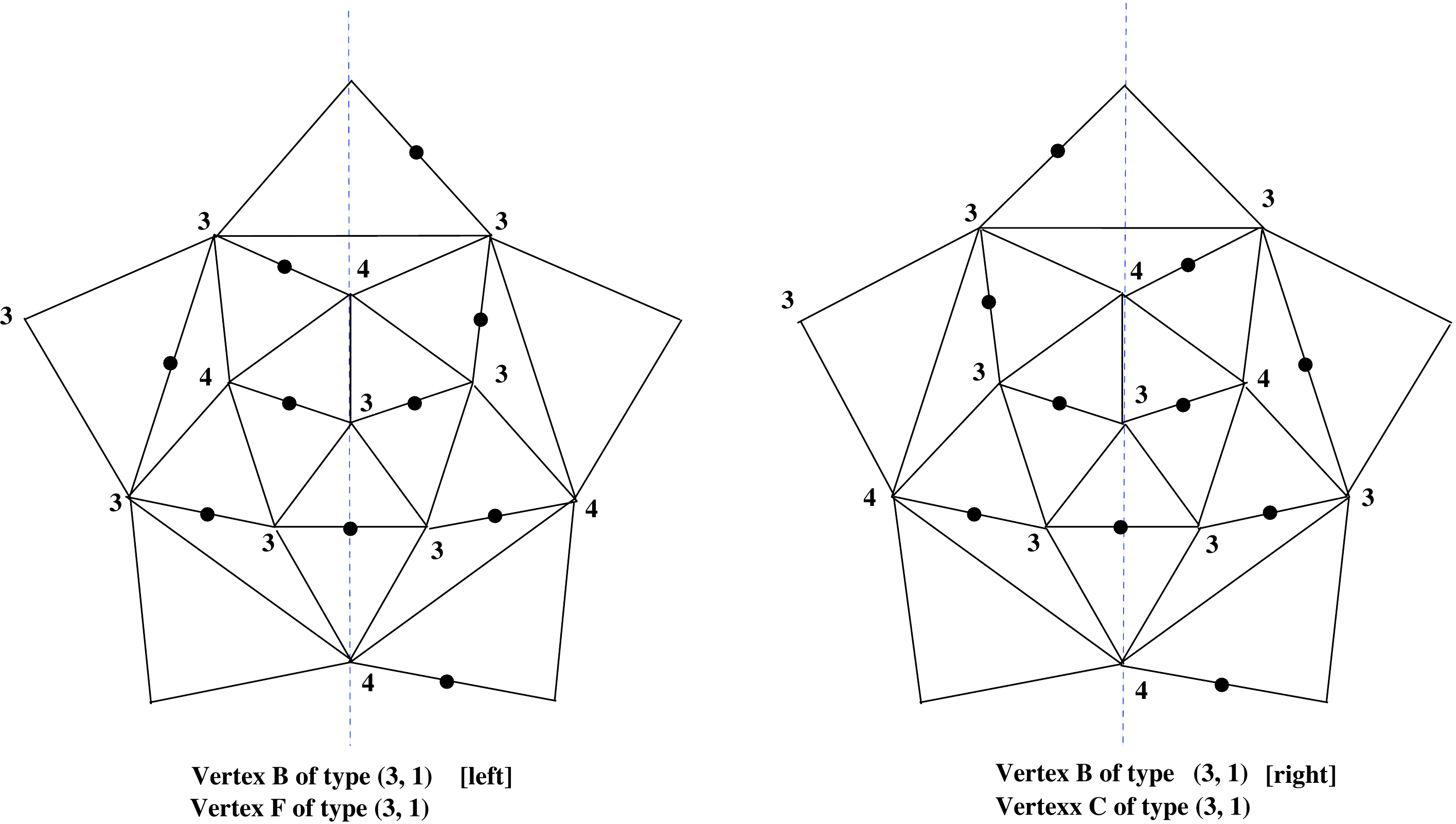}}
\end{center}
\caption{{\em Mirrored start configurations when the bead distribution $(1,3)$ is placed at vertex A, with either $(3, 1)_{\ell, B}$ and $(3, 1)_F$  (configuration 7 in Table~\ref{configs})), or with 
 $(3, 1)_{r, B}$ and $(3, 1)_C$ (configuration 9 in Table~\ref{configs})). The axis of mirror symmetry is the dashed blue line through vertices A, B, J and L.}}
\label{3b}\end{figure}

These considerations are summarized in Table \ref{configs}.  
\begin{table}[ht]
\begin{center}
\begin{tabular}{|c|c|c|c|c|c|c|c|c|} \hline
\multicolumn{9}{|c|}{{\em Systematics of minimal bead configurations}}\\ \hline
$(5, 0)_A$ & \multicolumn{3}{c| }{ $(3, 1)_A$} & \multicolumn{5}{c| } {$(1, 2)_A$}\\  \cline{5-9}
$(1, 2)_B$ &  \multicolumn{3}{c| }{ $(1, 2)_D$}&$(5, 0)_B$&\multicolumn{2}{c| }{ $(3, 1)_{\ell, B}$} &\multicolumn{2}{c| }{ $(3, 1)_{r, B}$} \\
$(1, 2)_C$ &  \multicolumn{3}{c| }{ $(1, 2)_E$}&$(1, 2)_C$&\multicolumn{2}{c| }{ $(1, 2)_C$} &\multicolumn{2}{c| }{ $(1, 2)_E$} \\  
$(1, 2)_D$ &  \multicolumn{3}{c| }{ $(1, 2)_J$}&$(1, 2)_D$&\multicolumn{2}{c| }{ $(1, 2)_D$} &\multicolumn{2}{c| }{ $(1, 2)_F$} \\ \cline{2-9}
$(1, 2)_E$ &$(3, 1)_B$&$(1, 2)_{\ell, B}$&$(1, 2)_{r, B}$&$(1, 2)_E$&$(1, 2)_F$&$(3, 1)_F$&$(1, 2)_L$&$(3, 1)_C$\\ 
$(1, 2)_F$ & X &$(1, 2)_C$&$(3, 1)_C$&$(1, 2)_F$&$(3, 1)_E$&$(1, 2)_E$&$(3, 1)_D$&$(1, 2)_D$\\ 
$(1, 2)_G$ & X &$(3, 1)_F$&$(1, 2)_F$&$(1, 2)_G$&$(1, 2)_G$&$(1, 2)_G$&$(3, 1)_G$&$(1, 2)_G$\\ 
$(1, 2)_H$ & X &$(1, 2)_G$&$(1, 2)_G$&$(1, 2)_H$&$(3, 1)_H$&$(1, 2)_H$&$(1, 2)_H$&$(1, 2)_H$\\ 
$(1, 2)_I$ & X &$(1, 2)_H$&$(1, 2)_H$&$(1, 2)_I$&$(1, 2)_I$&$(3, 1)_I$&$(1, 2)_I$&$(1, 2)_I$\\ 
$(1, 2)_J$ & X &$(3, 1)_I$&$(1, 2)_I$&$(5, 0)_J$&$(3, 1)_J$&$(3, 1)_J$&$(3, 1)_J$&$(3, 1)_J$\\ 
$(1, 2)_K$ & X &$(1, 2)_K$&$(3, 1)_K$&$(1, 2)_K$&$(1, 2)_K$&$(1, 2)_K$&$(1, 2)_K$&$(3, 1)_K$\\ 
$(5, 0)_L$ & X &$(3, 1)_L$&$(3, 1)_L$&$(1, 2)_L$&$(1, 2)_L$&$(1, 2)_L$&$(1, 2)_L$&$(1, 2)_L$\\ 
\hline
1&(2)&3&4&5&6&7&8&9\\ 
{ 4 strds}& & { 2 strds} &{ 2 strds}&{ 4 strds}& {2 strds} &{ 2 strds}& { 2 strds} &{ 2 strds}\\
\hline
\end{tabular}
\caption{{\em Summary of possible start configurations for the icosahedral cages prior to the identification of the  subset of symmetry-inequivalent configurations.}}
\label{configs}
\end{center}
\end{table}

The result is that, up to mirror symmetry, one has four inequivalent classes. One admits four strands (configurations 1 and 5 are equivalent under a rotation of the icosahedron), and the other three require two strands. Configurations 3 and 4 are mirror-symmetric, as are configurations 6 and 8, as well as 7 and 9. 

\begin{figure}[ht]
\begin{center}
\raisebox{-1.3cm}{\includegraphics[width=9.1cm,keepaspectratio]{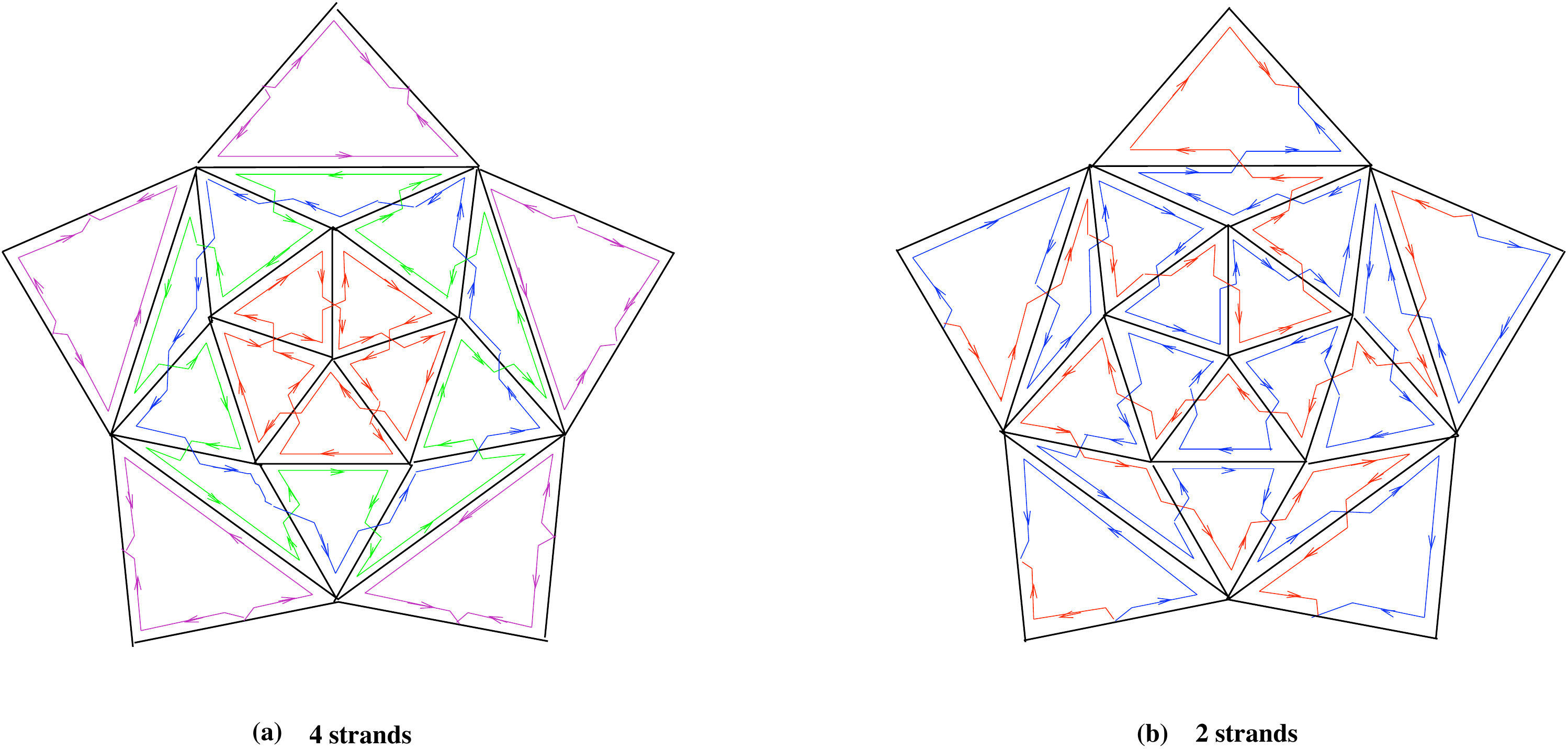}}
\end{center}
\caption{{\em (a) Four-strand start configuration 1, (b) two-strand start configuration 6.}}
\label{conf16}\end{figure}

\begin{figure}[ht]
\begin{center}
\includegraphics[width=5.0cm,keepaspectratio]{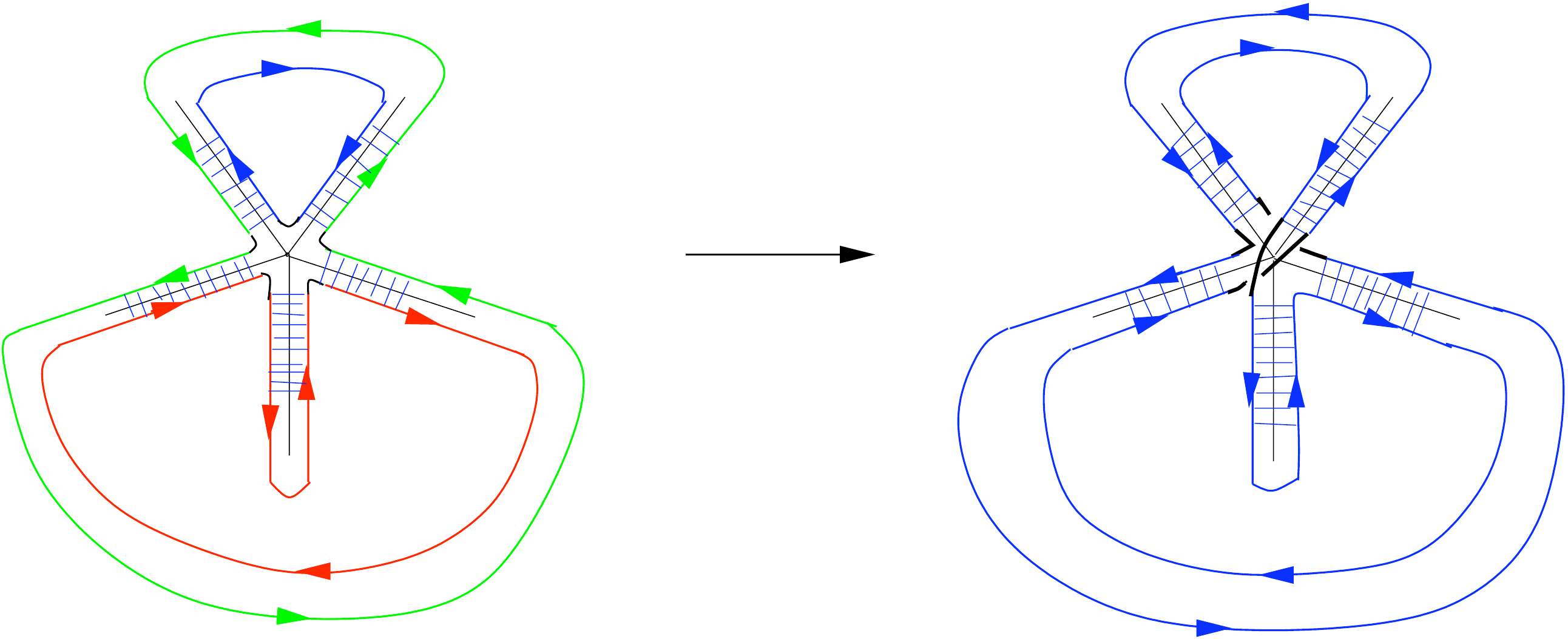}
\end{center}
\caption{{\em Replacement of a three-strand configuration by a one-strand configuration at a type B$_{2,5}$ junction.}}
\label{3strandjunction}
\end{figure}

The final step in the construction of such cages is to merge the different strands by using new types of junctions. The start configuration 1 corresponds to the four-strand cage of Fig.~\ref{conf16}(a). 
The figure shows that the two vertices (A and L, compare with the labelling in Fig.~\ref{vertices}(a)) host  thickened junctions of type A$_5$, containing five strands. All other vertices have thickened junctions of type $B_{2,5}$ , i.e., 5-junctions with two cross-overs of the strands at the vertices. These junctions involve three different strands, and via a replacement of strand connections as depicted in Fig.~\ref{3strandjunction} these strands can be merged into one. 
If such a replacement takes place at vertex B say (replacing the junction of type $B_{2,5}$), then the resulting cage is made of two distinct strands. The results in \cite{Jonoska} show that the parity of the boundary components in every thickened graph of a given polyhedron is always the same. Therefore, two strands are the least number of strands that can be used in the construction of an icosahedron. 

\begin{figure}[ht]
\begin{center}
(a)\,\,\includegraphics[width=4.9cm,keepaspectratio]{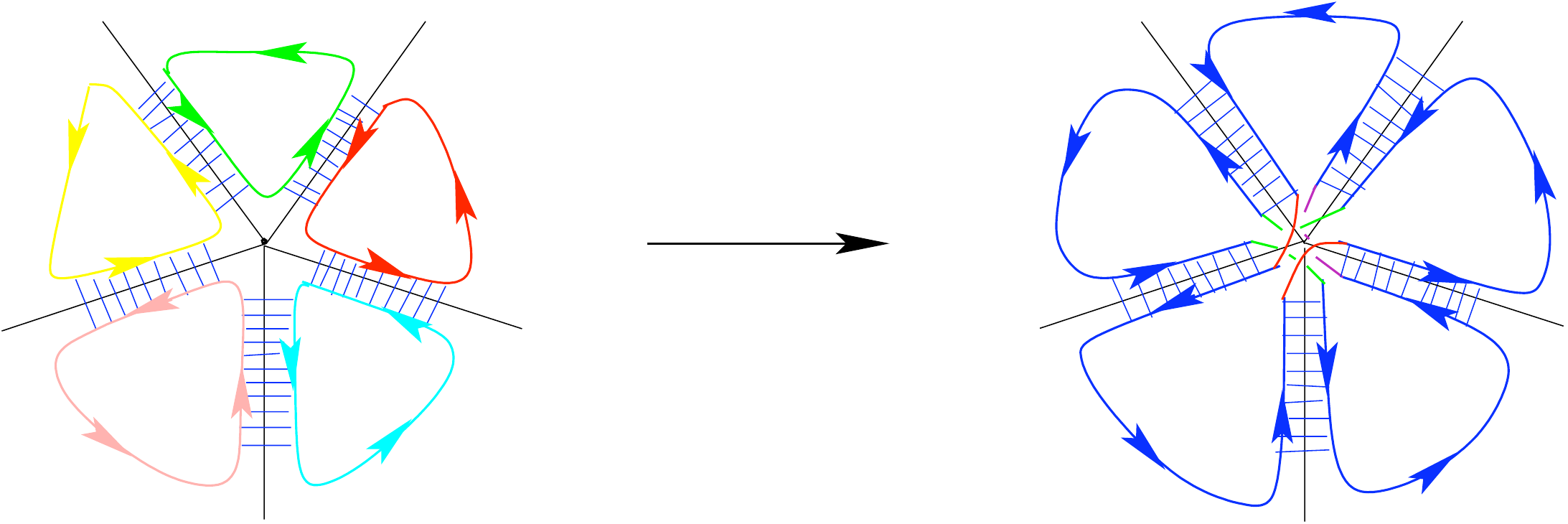}\qquad
(b)\,\,\includegraphics[width=4.9cm,keepaspectratio]{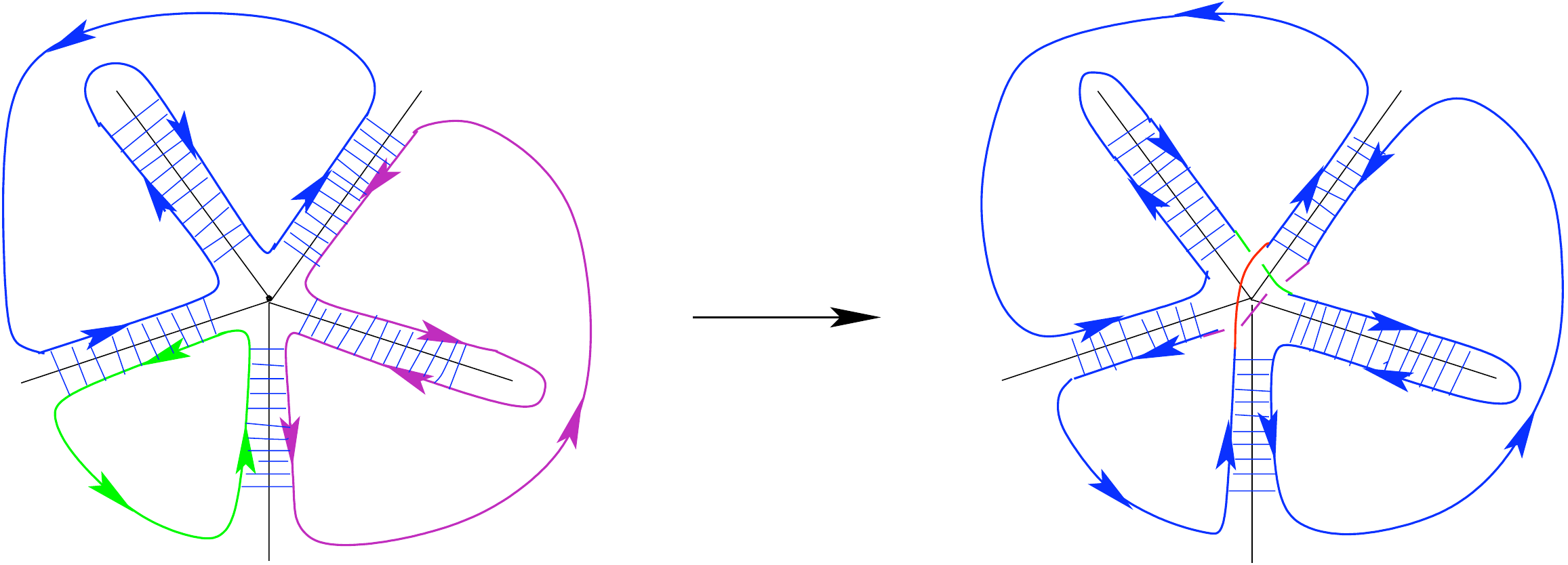}\\
\end{center}
\caption{{\em The junctions used as replacements for A$_5$ vertices.}}
\label{types}\end{figure}

For the case when an icosahedron has an edge length with an even number of helical half-turns, a start configuration is given by 20 strands of DNA tracing the polyhedral faces. These can be reduced to two strands in a number of ways. 
One possibility is to use the replacements shown in Fig.~\ref{types}, which result in (a) a reduction of five distinct strands meeting at a vertex to one, and (b) a reduction of three different loops to one. 
Via three replacements of type (a) and two of type (b) one obtains a two-strand realisation of the cage.

\section{The dodecahedral cage}\label{D}

\begin{figure}[ht]
\begin{center}
(a) \includegraphics[width=3.5cm,keepaspectratio]{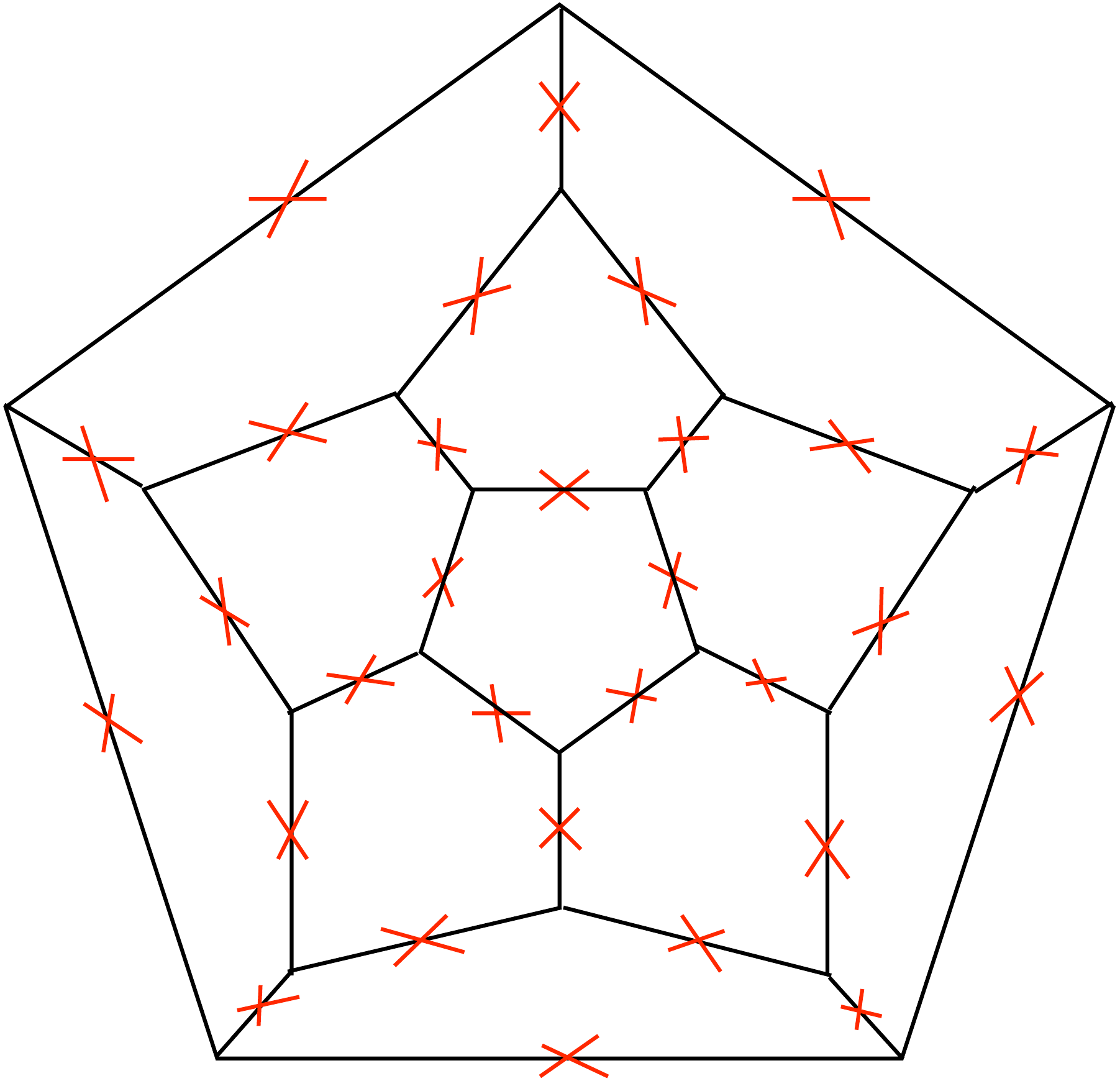}
(b) \includegraphics[width=2.0cm,keepaspectratio]{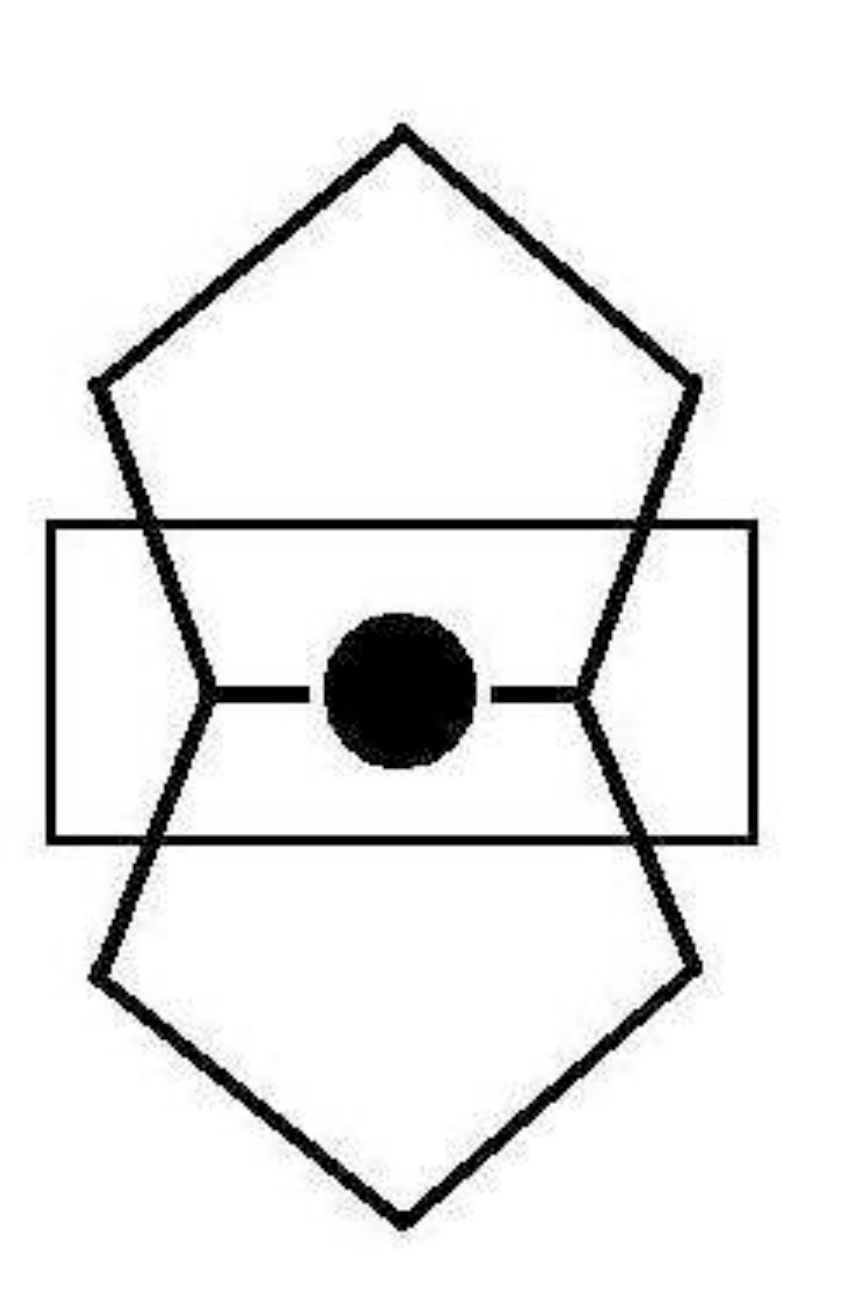}
\end{center}
\caption{{\em (a) A dodecahedral cage in planar projection, with 
cross-over schematically representing the twists that occur along the edges as a result of an odd number of half-turns in the helical structure. (b) Tiles needed to make the structure orientable.}}
\label{Fig3}
\end{figure}

Next we consider the case of the dodecahedron, a three-coordinated polyhedron with 12 pentagonal faces. We start with the case where the edge length of the dodecahedral cage is such that the DNA duplex has an odd number of half-turns along each edge, and we encode this additional helical turn via a red cross-over on a dodecahedral cage in planar projection (see Fig.~\ref{Fig3}(a)). 
Since the resulting number of twists along a pentagonal face is odd, the surface corresponding to the ribbon representing the double helical structure of the DNA is non-orientable. We follow the procedure outlined in Section \ref{sec1} to produce orientable embeddings that may serve as start configurations for our analysis. In particular, we replace $A_3$ junctions by $B_{1,3}$ junctions in order to compensate for the twist, and determine their locations via the bead rule. 

\begin{figure}[ht]
\begin{center}
\includegraphics[width=5cm]{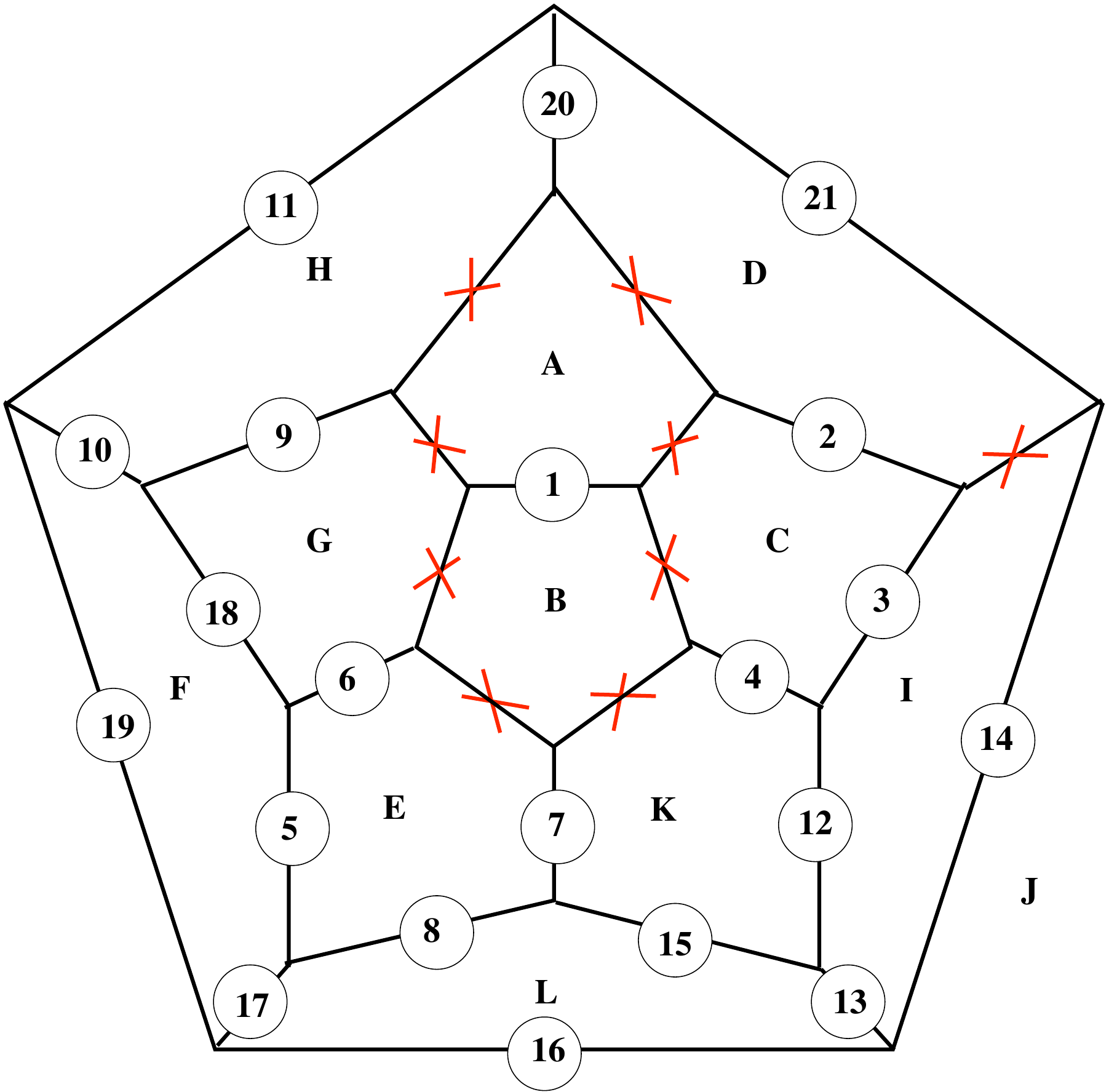}
\end{center}
\caption{{\em Labelling system corresponding to Fig. \ref{B}. Symbols (letters, numbers) identifying the 
 pentagons follow the  first branch of the solution tree in Fig. \ref{B}.}}
\label{A}
\end{figure}

As there are twelve faces in the dodecahedron, the minimal number of beads needed to provide an orientable embedding of the whole double helical structure is six. There are several non-equivalent ways to introduce six beads by tessellating the surface of the dodecahedron with the tile shown in Fig.~\ref{Fig3}(b). The inequivalent possibilities of placing beads on the edges of the dodecahedron such that precisely one edge per pentagonal face is decorated by a bead are obtained as follows. Without loss of generality, we place the first bead on any of the edges, because they are all equivalent by symmetry. We label this edge, and hence the bead on this edge, as 1; see Fig.~\ref{A}. 
To keep the number of beads at a minimum, no other edge of the two pentagons labelled A and B can have a bead. 

\begin{figure}[ht]
\begin{center}
\includegraphics[width=5.8cm]{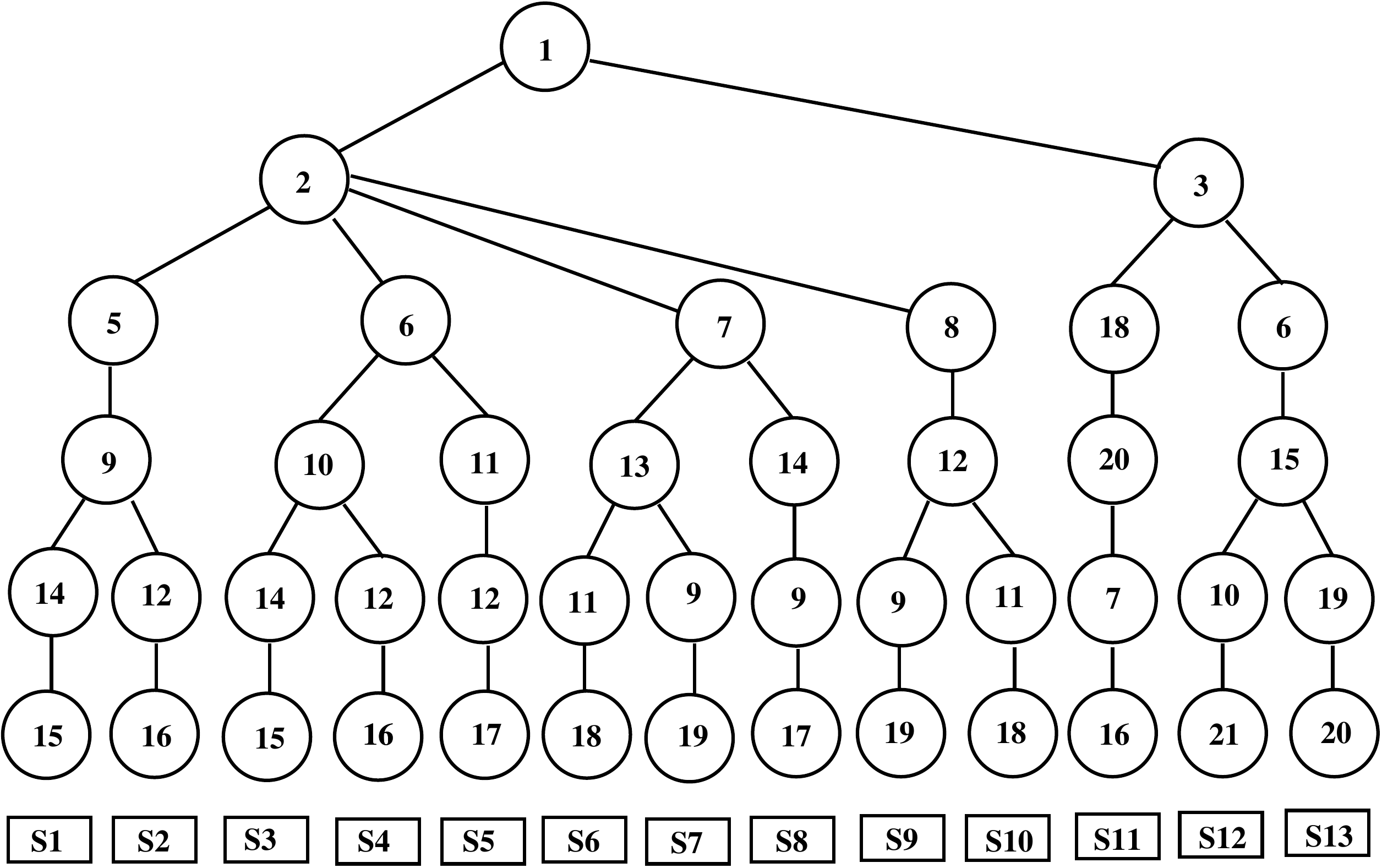}
\end{center}
\caption{{\em The tree encoding the 13 solutions of placing six beads on the dodecahedral edges such that each pentagon has precisely one bead. Only solutions S1, S2, S3, S6 and S11 are inequivalent.}}
\label{B}
\end{figure}

Since the order in which pentagons are considered does not matter, we can next concentrate on the pentagon labelled C. Edges labelled 2 and 4 are equivalent by symmetry and hence lead to equivalent configurations. We therefore consider only the possibility of placing a bead on edge 2 or edge 3, and treat each case separately. A bead on edge 2 `covers' pentagons C and D in Fig. \ref{A}. 
Since every pentagonal face has to have a bead on one of its edges, we can, without loss of generality, look at pentagon E next. Due to the beads already present there is no symmetry, and thus all edges 5 to 8 lead to different solutions. 
There are 10 solutions in total, denoted S1 to S10 in Fig. \ref{B}. 
We next concentrate on edge 3 instead of edge 2, hence pentagons A, B, C and I contain an edge with a bead.
Up to symmetry arguments,  there are three different ways to complete this analysis based on the bead rule, labelled S11 to S13 in Fig. \ref{B}. 

We investigate whether any of these 13 solutions can be symmetrically mapped onto another. 
To do so, we observe that different bead configurations can be classified according to  
different {\it bead paths}: 
 We call any path along edges of the decorated dodecahedron a  bead path if, starting and ending on a beaded edge, it  
 alternates along non-beaded and beaded edges. 
The four possible scenarios of bead paths are summarised in Table \ref{configs2}. 

\begin{table}[ht]
\begin{center}
\begin{tabular}{|c|l|} \hline
\multicolumn{2}{|c|}{{\em Systematics of bead paths}}\\ \hline
Type I & all beads lie on a closed (cyclic) bead path \\
& (corresponding to solutions S1, S2, S4, S9) \\
Type II &  all beads lie on a non-cyclic (open) bead path  \\
& (corresponding to solutions S3, S5, S7, S8, S10, S12)\\  
Type III &  all beads lie on three disjoint bead paths  
with two beads each\\ & (corresponding to solutions S6, S13) \\ 
Type IV & all beads lie on six  disjoint bead paths one edge in length   \\
 & (corresponding to solution S11)\\ 
\hline
\end{tabular}
\caption{{\em Summary of possible bead paths for the  identification of the  symmetry-inequivalent start configurations.}}
\label{configs2}
\end{center}
\end{table}

Bead placements corresponding to a given type (I--IV) are inequivalent to arrangements  
of any other type, because any symmetry would map a path onto another path, and a cycle onto another cycle.
Hence an equivalence may appear only within a given type. 
We start with type I. Solutions S1 and S9 are mirror images of each other (obtained by turning the sphere inside out, i.e.,  they have different helical structures but are otherwise identical). We therefore consider them as equivalent. Solutions S2 and S4 can be considered equivalent as they correspond to the closed path 
that divides the 12 pentagons of the dodecahedron into two identical sets of six. However, S1 and S2 are not related by symmetry, and there are hence two distinct solutions of type I, represented by S1 and S2. Solutions S3, S5 and S8 represent
mutually   symmetric open paths, while S7, S10 and S12 represent mutually   
 symmetric  paths, but   of opposite orientation  to S3, S5, and S8  
   (again obtained by turning the sphere inside out). Hence, we can  
   consider all these  
   options as equivalent, and so there is only one 
   (up to symmetry) arrangement of beads producing a non-cyclic   
   path, represented by S3. The two  arrangements  
   of type III corresponding to solutions S6 and S13 can be mapped one onto another.  So there is only one solution of this type, represented, say by S6. Finally, S11 is the only solution of type IV. It corresponds to an equidistant distribution of beads on edges and is in that sense the `most symmetric' solution.

\begin{figure}[]
\begin{center}
(a) \includegraphics[width=4.6cm,keepaspectratio]{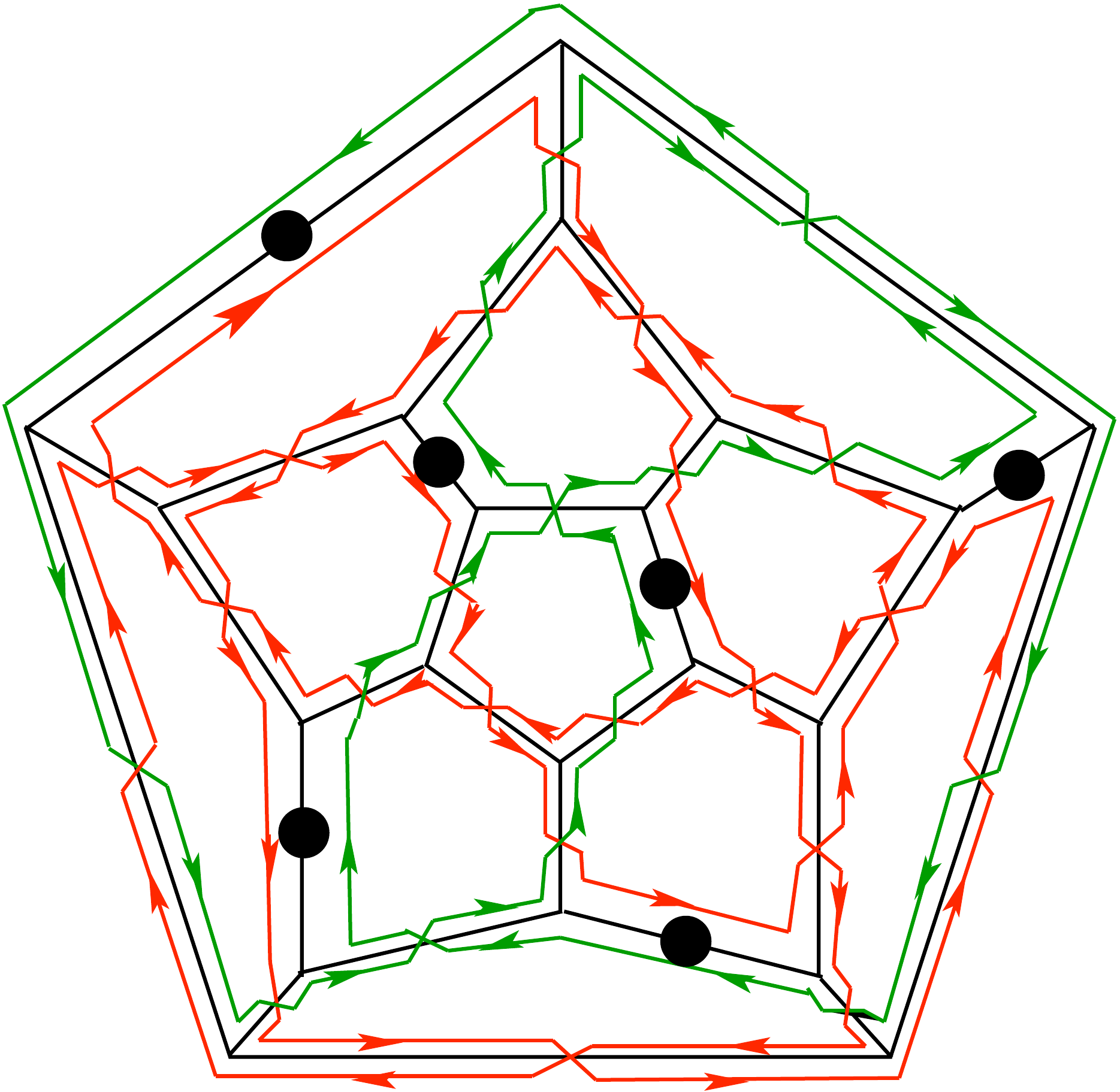}
(b) \includegraphics[width=4.6cm,keepaspectratio]{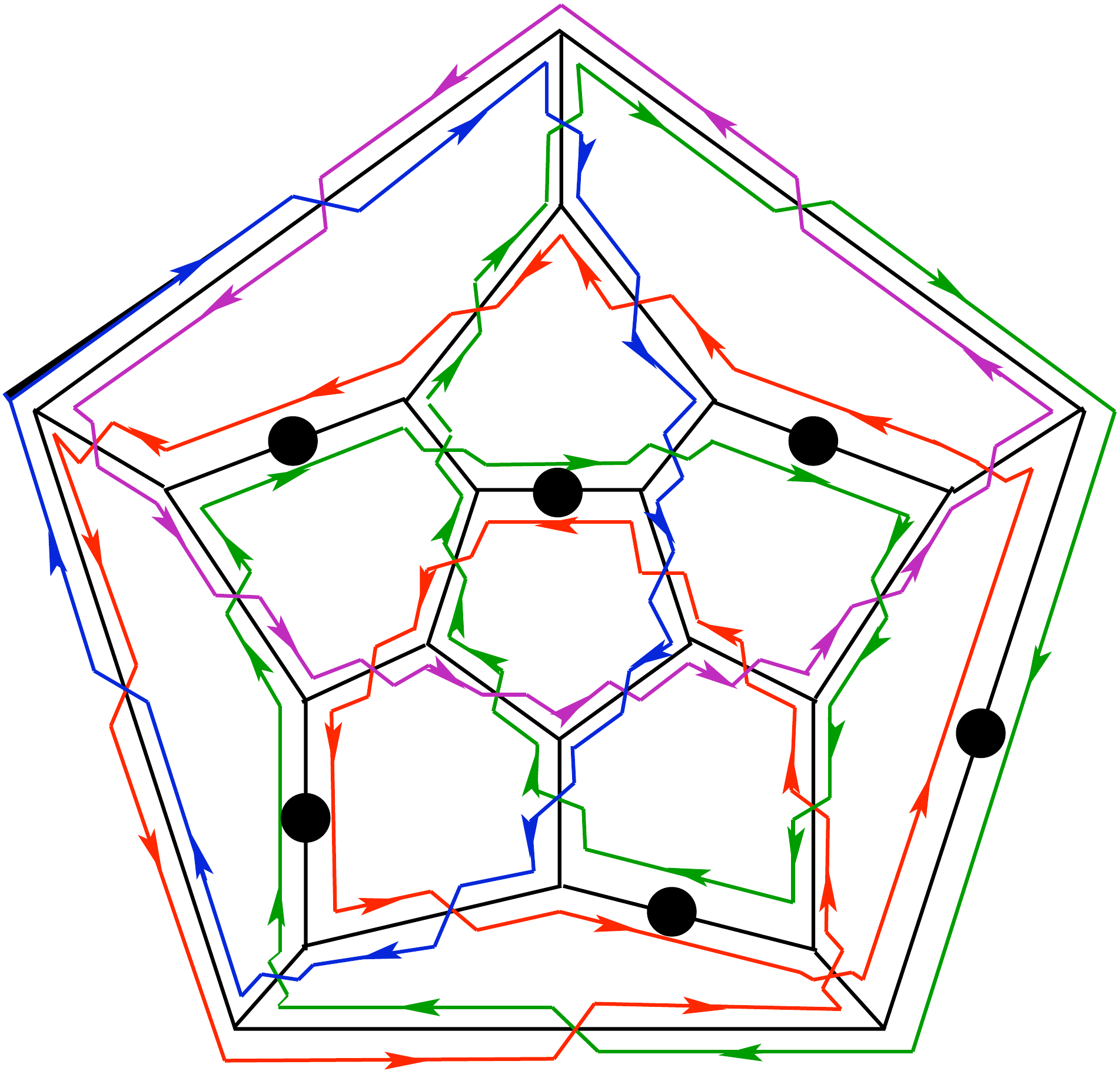}\\
(c) \includegraphics[width=4.6cm,keepaspectratio]{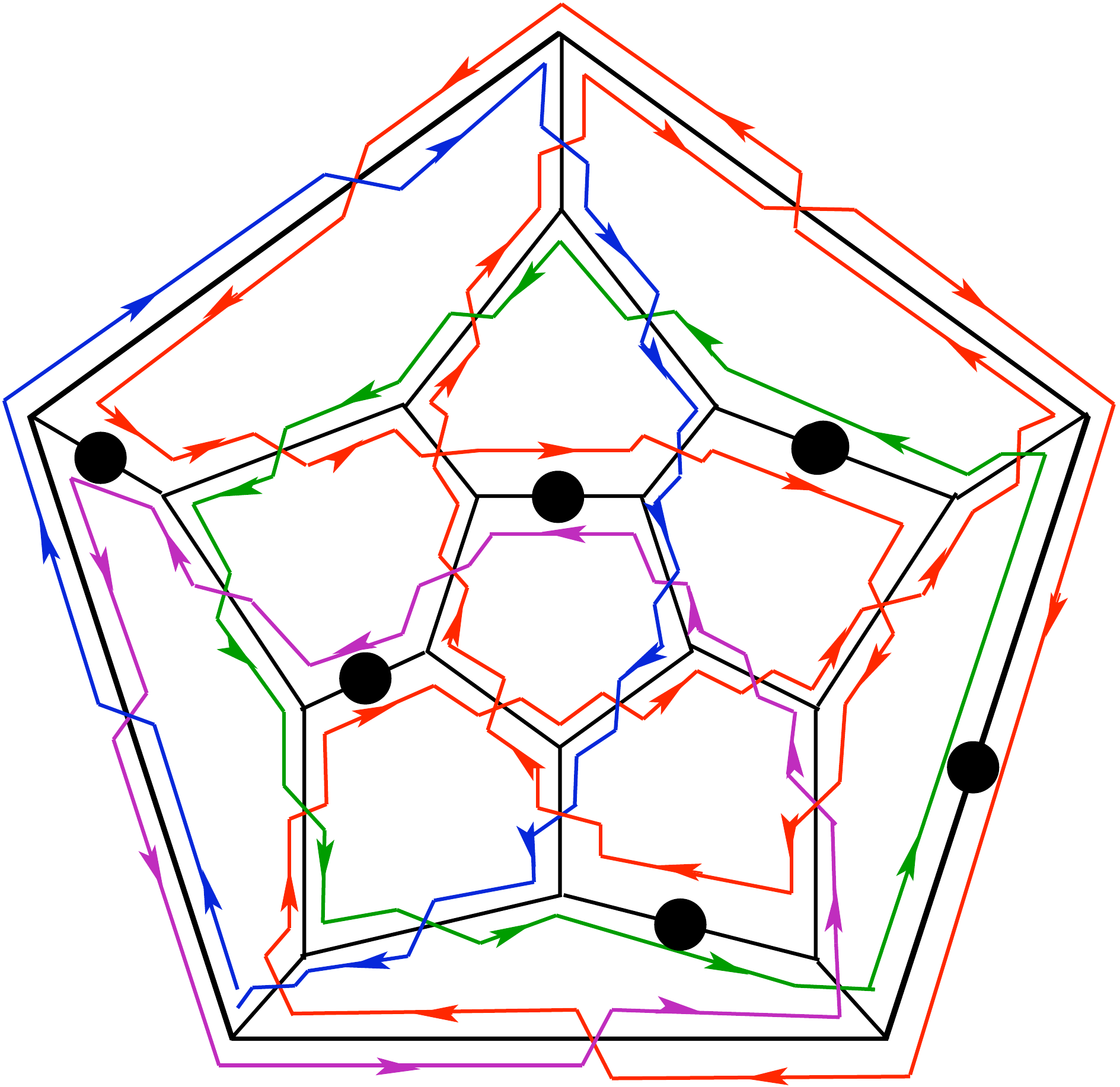}
(d) \includegraphics[width=4.6cm,keepaspectratio]{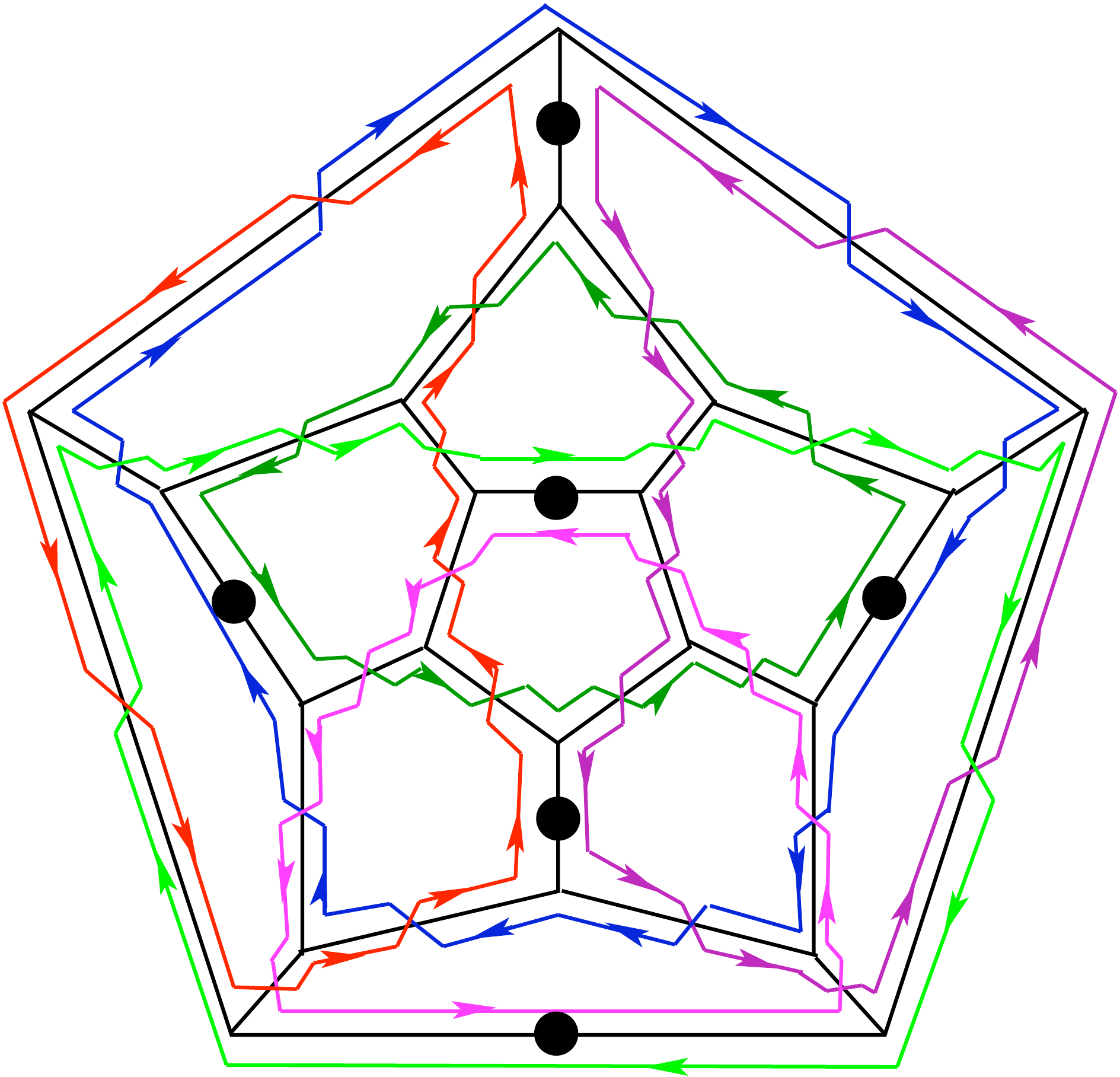}\\
(e) \includegraphics[width=4.6cm,keepaspectratio]{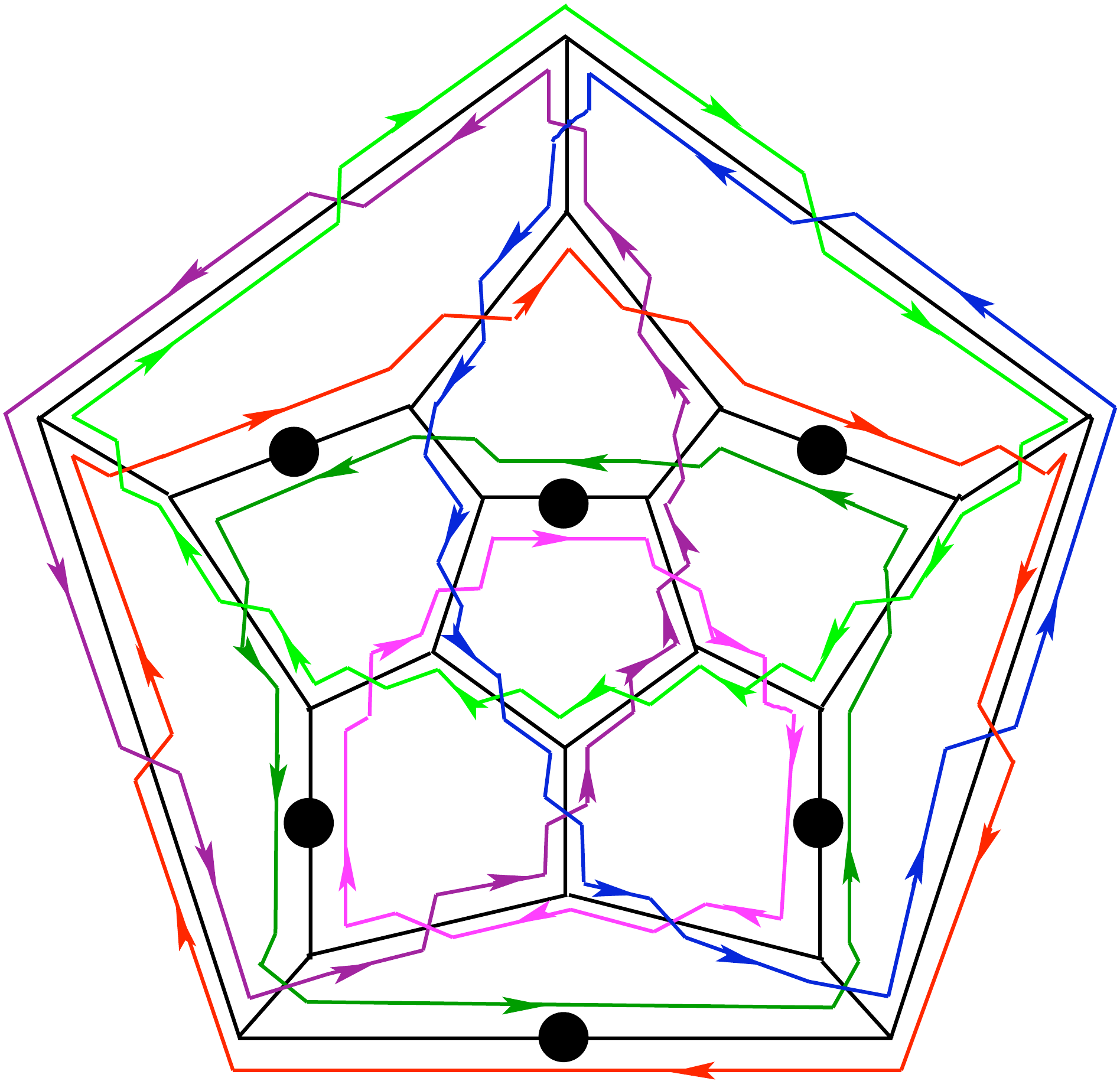}
\end{center}
\caption{{\em Embeddings of the DNA duplex structures in a dodecahedral cage corresponding to the inequivalent bead configurations. The bullets indicate the placements of beads, i.e., the edge containing an even number of half-turns.
The configurations correspond to (a) solution S6 with two separate strands;  (b) solution  S1 with four separate strands;  (c) solution S3 with four separate strands; (d) solution  S11 with six separate strands; and (e) solution  S2 with six separate strands.}}
\label{strand-configs}
\end{figure}

Every bead configuration translates into a start configuration. The five inequivalent start configurations
are depicted in Fig.~\ref{strand-configs}. 
Each of these DNA embeddings  has 14 vertex configurations of type $A_3$ and 6 vertex configurations of type $B_{1,3}$. 
The DNA embeddings corresponding to solutions S2 and S11 result in six distinct strands (Fig. \ref{strand-configs}(d,e)), the embeddings corresponding to solutions S1 and S3 in four strands (Fig. \ref{strand-configs}(b,c)), while the embedding corresponding to solution S6 in two strands (Fig. \ref{strand-configs}(a)).

\begin{figure}[]
\begin{center}
\includegraphics[width=4.3cm,keepaspectratio]{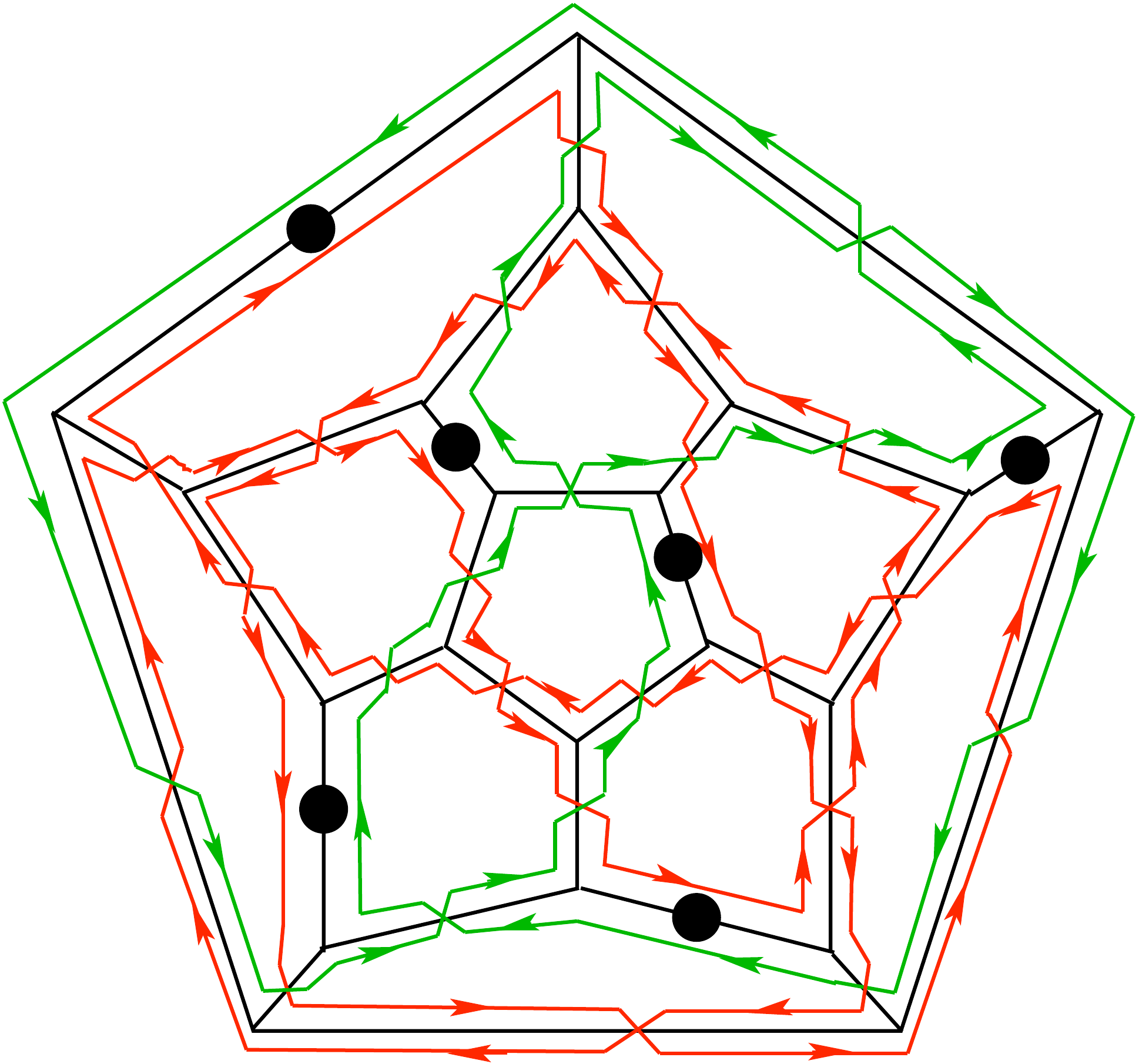} 
\end{center}
\caption{{\em Embedding of a duplex  DNA structure corresponding to the solution S6 with two strands.}}
\label{Fig5}
\end{figure}

In order to realise the cage in terms of a minimal number of DNA strands, the configurations in 
 Fig.~\ref{strand-configs}(b--d) require further vertex replacements. Note that as in the case of the icosahedron, (by the results in \cite{Jonoska}) two is the least number of strands that can be used to obtain the dodecahedral cage. Indeed, if a vertex 
 configuration    of type $A_3$  obtained by hybridizing three separate strands is replaced by a vertex configuration of a branch point of type $B_{1,3}$, then the number of strands meeting at the vertex reduces to one according to \cite{Jonoska}. In the case of six separate strands (i.e.,  the cases of S2 and S11) we need to  
choose two vertices for replacement, while for the case of four separate strands (i.e.,  the cases of S1 and S3) only one replacement is needed. The reader is referred to \cite{JT} for details. In all cases, the minimal number of strands to assemble a dodecahedron using only junctions of type $A_3$ and $B_{1,3}$ is two.  Fig.~\ref{Fig5}(a) shows  the two-strand configuration corresponding to solution S6. 

We now consider the case  of double-stranded DNA embeddings into dodecahedral  cages with {\it an even number of half-turns per edge}. In these cases, there are no additional twists on the edges. A start configuration is therefore easily obtained by placing 12 separate strands into the 12 faces of the dodecahedron. Via $n=5$ replacements, one again obtains a  DNA embedding with two separate strands. According to the remark at the end of Section \ref{sec1} (see also details in \cite{JT}), these replacements can be freely chosen if flexible strand connections are used, and require particular positions in the non-flexible case. 

\section{The icosidodecahedral cage} \label{IDD}

We now consider the icosidodecahedron, a 32-faced polyhedron with quatro-valent vertices. We again start with the case where all edges accommodate an odd number of helical half-turns, which implies additional cross-overs on the edges in the initial configuration. According to Section \ref{sec1}, the bead rule has to be applied in order to obtain a start configuration. The minimal number of beads required is easily calculated. All faces of the icosidodecahedron have an odd number of edges: there are 12 pentagons and 20 triangles. However, each face must have an even number of cross-overs to keep the orientability. Therefore, each of the 20 triangles must receive at least one bead. However, by placing a bead on each triangle, one also places at least one bead on each pentagon. The distribution of this minimum number of  20 beads should be such that pentagons receive an odd number of beads. Let $\alpha$ be the number of pentagons receiving one bead, $\beta$ be the number of pentagons receiving three beads and $\gamma$ be the number of pentagons receiving five beads. Given that there are 12 pentagons in total, we must satisfy the two equations
\begin{equation}
\begin{array}{rcl} 
\alpha+3\beta+5\gamma&=&20,\nonumber\\
\alpha+\beta+\gamma&=&12,
\end{array}
\end{equation}
with $\alpha, \beta$ and $\gamma$ positive integers. There are three solutions to the problem, namely
\begin{equation} \label{3cases}
\begin{array}{rcl}
&&{\rm Case\, I}\qquad\qquad \,\,\,\,\alpha=8, \, \beta=4,\, \gamma=0\nonumber\\
&&{\rm Case\, II}\qquad \qquad\,\,\alpha=9, \, \beta=2,  \, \gamma=1\nonumber\\
&&{\rm Case\, III}\qquad \qquad\alpha=10,  \, \beta=0,  \, \gamma=2.
\end{array}
\end{equation}
We start by considering case I. This tells us that the bead rule is fulfilled if there are four pentagons with three beads each, and if every triangle has precisely one bead. We therefore determine all symmetry-inequivalent start configurations with that property. Since this is a significant combinatorial task for the polyhedron under consideration, our analysis is computer-assisted. 

In a first instance, we use the icosahedral symmetry to reduce the number of options to be considered. In particular, we determine all symmetry-inequivalent  distributions of four pentagonal faces on the icosidodecahedron. Each of these four faces has three of its edges decorated by one bead, whilst all other pentagons and all triangles have only one bead on their perimeter.
In order to determine all inequivalent configurations of four pentagons, we consider the equivalent problem of finding all different possibilities of colouring four of the 12 vertices of an icosahedron.  

\begin{figure}[]
\begin{center}
(a)\includegraphics[width=2.5cm,keepaspectratio]{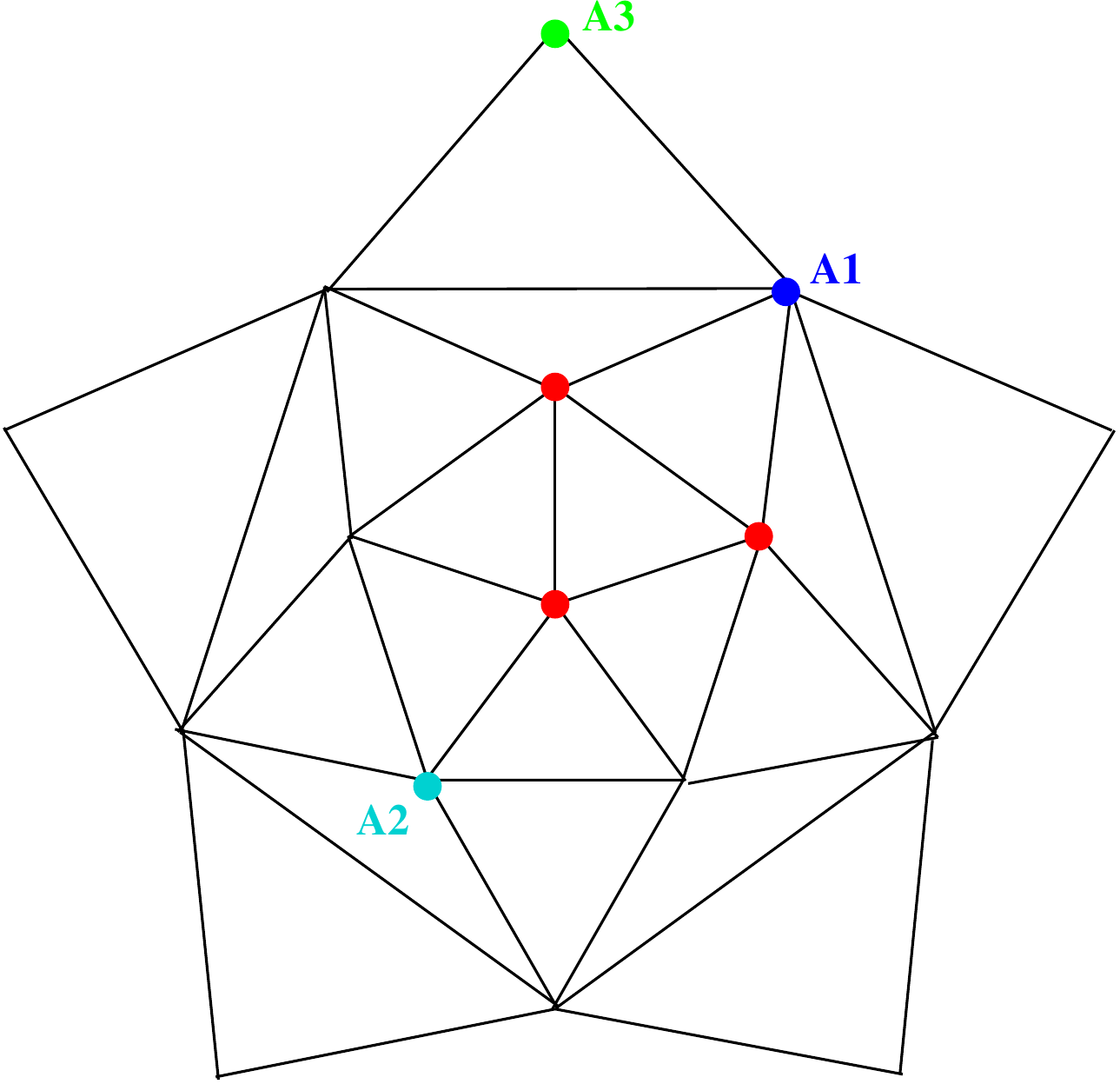}
(b)\includegraphics[width=2.5cm,keepaspectratio]{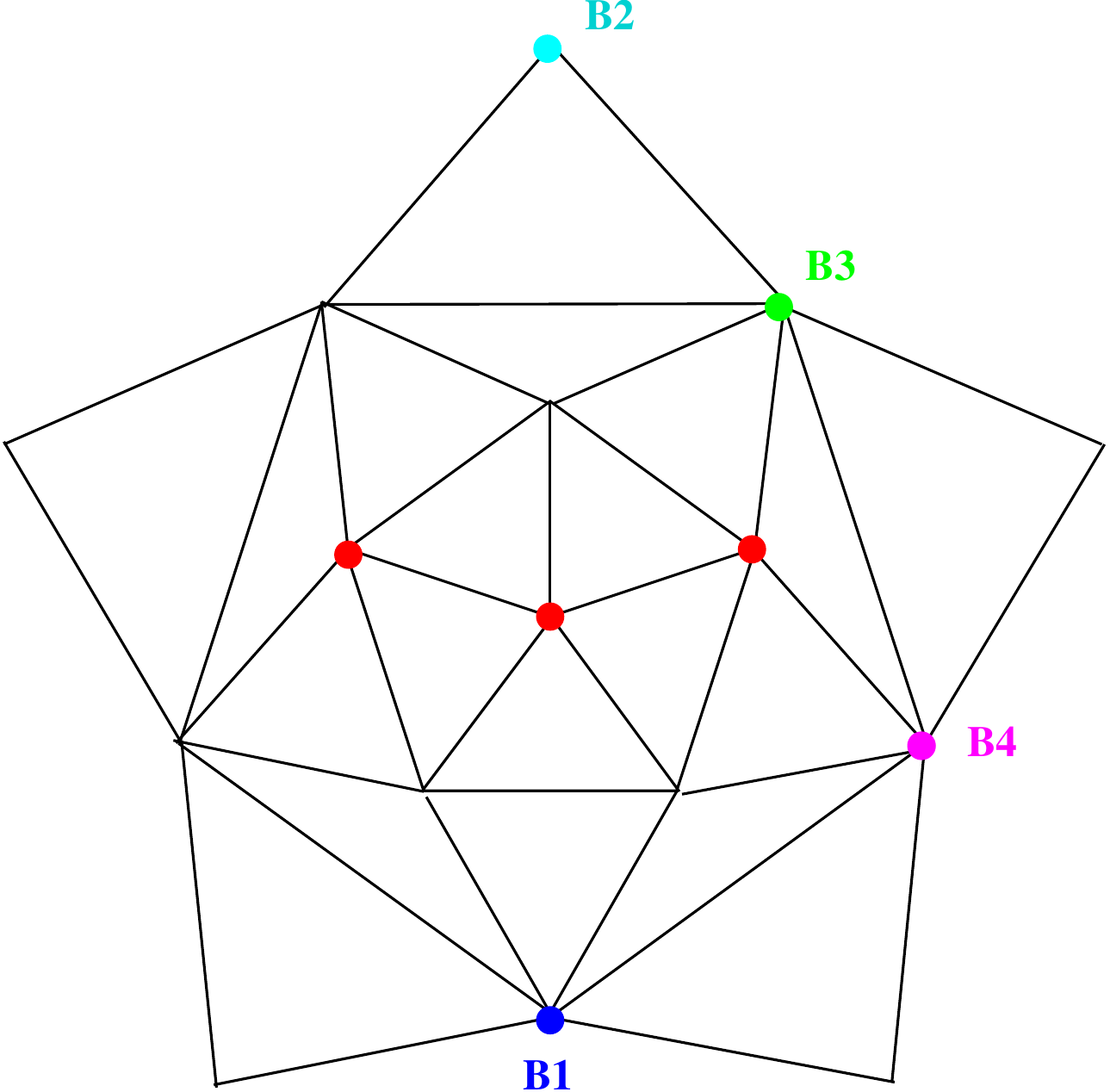}
(c)\includegraphics[width=2.5cm,keepaspectratio]{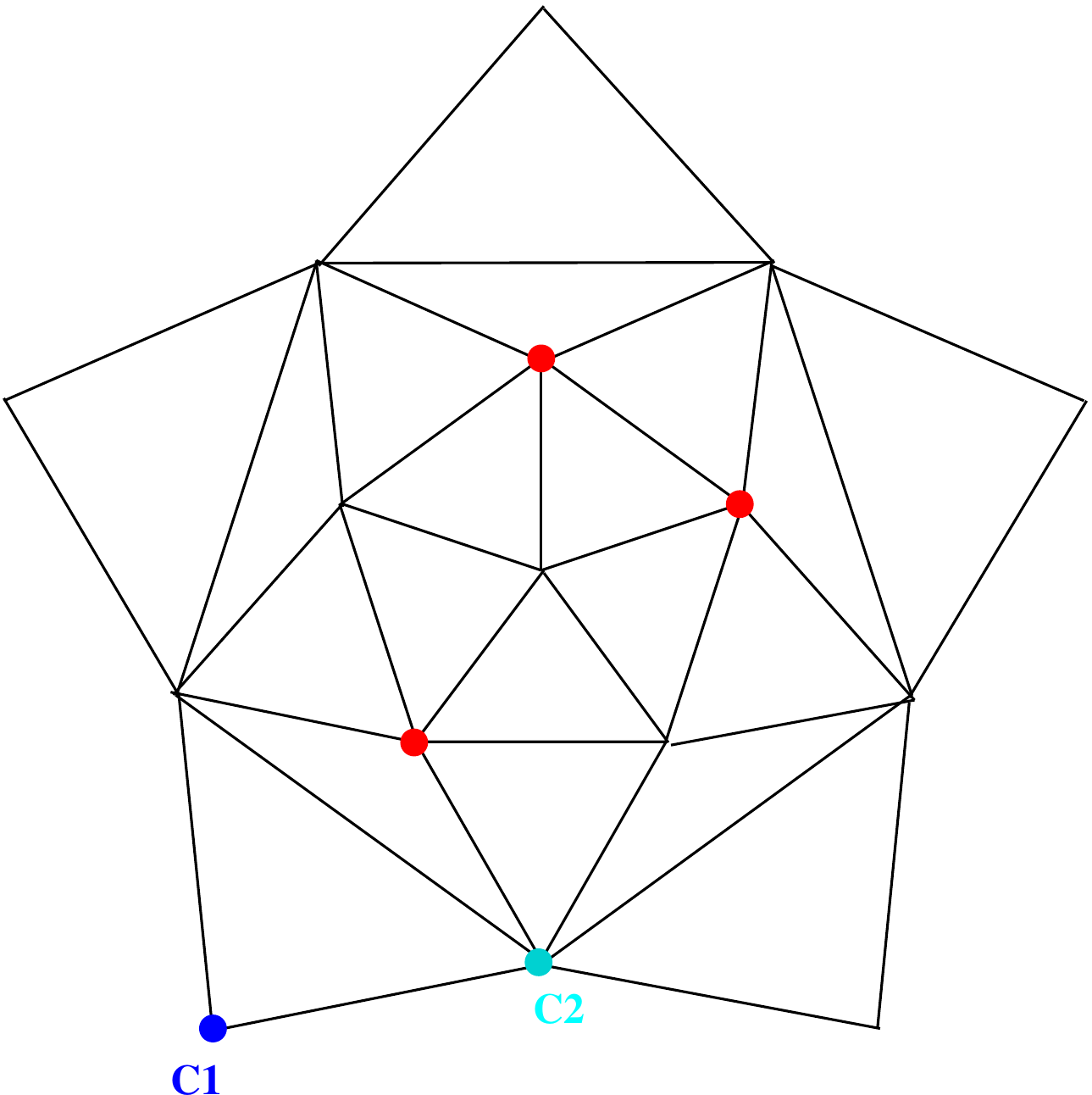}
\end{center}
\caption{{\em Partial start configurations for case I, with a distribution of four out of twelve pentagons on the icosidodecahedron, represented here as distributions of four vertices  on an icosahedron (a) configurations with three vertices being those of a triangle (red) and the fourth vertex being either $A_1$, $A_2$ or $A_3$, (b) configurations with three red vertices and the fourth one being either $B_1$, $B_2$, $B_3$  or $B_4$, (c) configurations with three red vertices and the fourth one being either $C_1$ or $C_2$.}}
\label{partialstart}
\end{figure}

There are 9 inequivalent such configurations for case I, which we call the {\it partial start configurations}. We show them schematically in a projective view of the icosahedron in Fig.~\ref{partialstart}. 

We next determine the inequivalent bead configurations for each partial start configuration.  
Each partial start configuration encodes several possible cage scenarios which correspond to all inequivalent ways of placing three beads on the edges of four distinguished  pentagons, and one bead on one of the edges of all other faces (pentagonal or triagonal). We carry out this combinatorial task computationally via a computer programme that tests, for each start configuration, which combinations of beads are possible, given the fact that the four distinguished pentagonal faces each have three beads, while all others have one. The results for all three cases are summarised in Table \ref{tab1}. 
\begin{table}[h]
\centering
\begin{tabular}{|c | c | c | c |}
\hline
Loop number & Case I & Case II & Case III \\ \hline
10 & X & & \\
12 & X & & \\
14 & X & X & X \\ 
16 &  &  & X \\ \hline
 \end{tabular}
\caption{\em The distribution of 10-, 12-, 14- and 16-loop configurations among the three different cases. A cross indicating the occurrence of this type of loop configuration for a given case.}
\label{tab1}
\end{table}

\begin{figure}[]
\begin{center}
(a)\,\,\includegraphics[width=4.5cm,keepaspectratio]{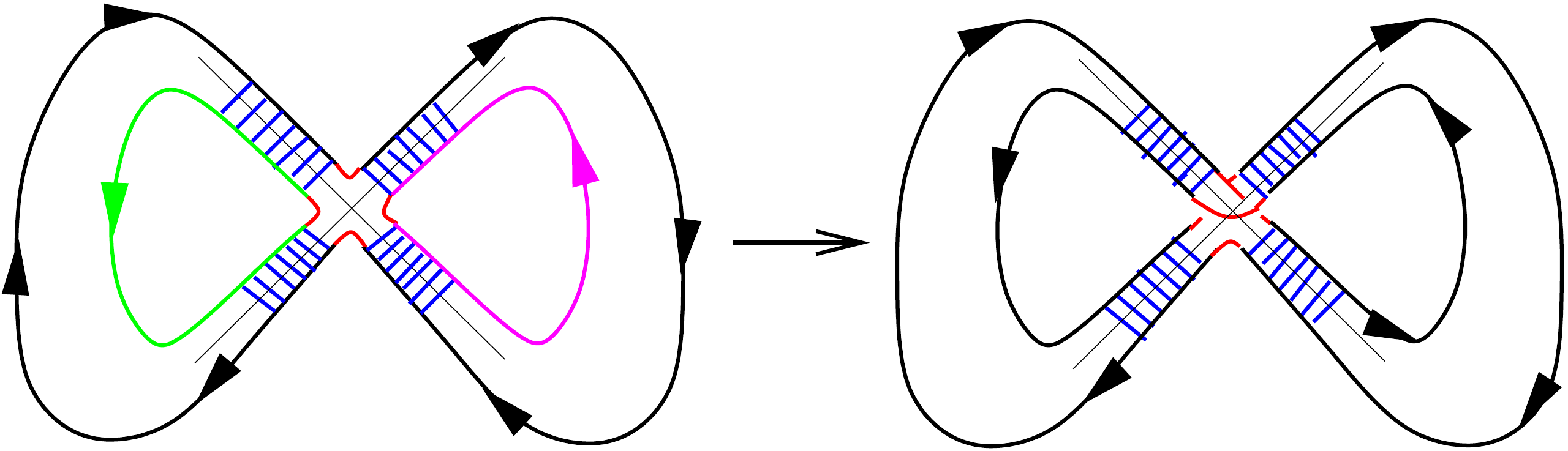}\qquad
(b)\,\,\includegraphics[width=4.5cm,keepaspectratio]{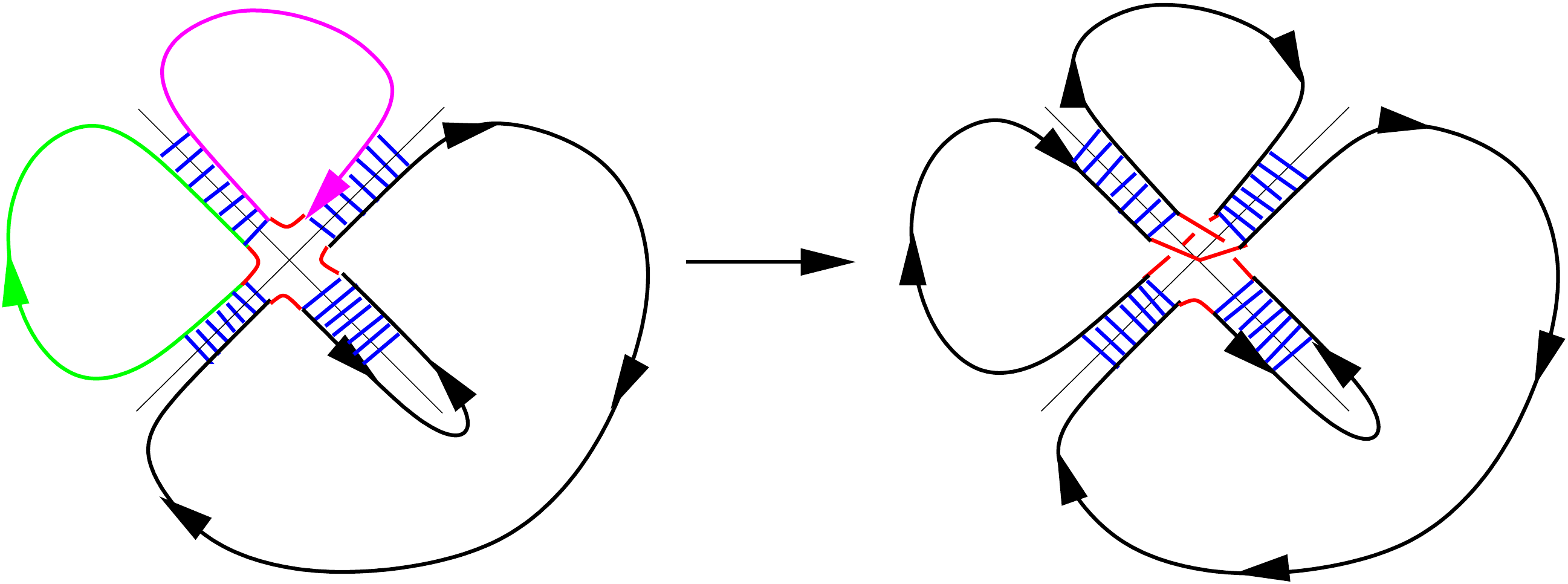}
\end{center}
\caption{{\em Replacements of  a 3-loop configuration  by a
single loop: (a) type I, (b) type II.}}
\label{replace}
\end{figure}

The computations show that the smallest number of distinct loops is 10 and the largest number 16. While there are over $10^4$ distinct 10-loop configurations, there is only one configuration with 16 loops, occurring in case III. All vertex configurations involve either four, three, or two distinct loops. As before, these can at best be reduced to two-strand configurations because the number of distinct circular DNA strands is even in all cases. Such a reduction is possible via the replacements of type I and type II shown in Fig.~\ref{replace}. 

If the edge length of the icosidodecahedron is such that none of the edges has an additional cross-over, the start configuration consists of 32 loops corresponding to the 32 faces of the polyhedron. Via 30 replacements of type I or type II, a duplex cage structure realised by two circular DNA molecules is obtained. 

\section{Discussion}

We have performed a theoretical analysis of different types of icosahedral cages with minimal (two) DNA strands such that each edge is in a duplex structure. Focus was placed on the three icosahedrally symmetric polyhedra with the smallest number of vertices, the icosahedron with 12 vertices at the 5-fold axes of icosahedral symmetry, the dodecahedron with 20 vertices at the 3-fold axes, and the icosidodecahedron with 30 vertices at the 2-fold axes. These polyhedra are distinguished by the fact that they have uniform edge lengths and may therefore be easier to manufacture than other polyhedra with icosahedral symmetry. 

We remark that polyhedral RNA cages have been observed also within the protein containers, called viral capsids, that encapsulate and hence provide protection for the viral genome \cite{Tang,Worm}. However, the RNA has to be unknotted for successful replication and therefore these cages usually do not appear in the form we have described here and are realised by the viral RNA in a different way \cite{Ranson,Bruinsma}. 

A comparative analysis reveals interesting features of the cage structures. In particular, for each of them the minimal number of circular molecules needed to construct the cage is two. The icosidodecahedral cage is distinguished by the fact that its volume per surface ratio is the largest among the polyhedra considered here (and is also larger than those of the cages that have been realised experimentally to date). It may  therefore be more suitable for nanotechnology applications in which the cages serve as containers for storage or the transport of a cargo. 

We hope that the blueprints for the organisation of the cages with a non-crystallographic symmetry suggested here may assist in their experimental realisation. In particular, these blueprints suggest the structures of the junction molecules that may be used as basic building blocks for the self-assembly of those cages along the lines of \cite{JACS,AJS}.

\section*{Acknowledgements}
AT has been supported by an EPSRC Springboard Fellowship, and RT by an EPSRC Advanced Research Fellowship. Some of this work has been carried out at the University of South Florida, and both AT and RT would like to thank the Department of Mathematics for their hospitality. NJ has been supported in part by the NSF grants CCF $\#$0523928
and CCF $\#$0726396.


\end{document}